\documentclass[12pt,preprint]{aastex}

\begin{document}

   \title{Nuclear activity and massive star formation in the low luminosity AGN 
     NGC~4303: \emph{Chandra} X-ray observations}

\author{E. Jim\'enez-Bail\'on}
\affil{LAEFF-INTA, POB 50727, E-28080 Madrid, Spain}
\author{M. Santos-Lle\'o}
\affil{XMM-Newton Science Operations Center, VILSPA, ESA, POB 50727,
E-28080 Madrid, Spain}
\author{J.M. Mas-Hesse}
\affil{CAB(CSIC-INTA), E-28850 Torrej\'on de Ardoz, Madrid, Spain}
\affil{LAEFF-INTA, POB 50727, E-28080 Madrid, Spain}
\author{M. Guainazzi}
\affil{XMM-Newton Science Operations Center, VILSPA, ESA, POB 50727,
E-28080 Madrid, Spain}
\author{L. Colina}
\affil{Instituto de Estructura de la Materia (CSIC), Serrano 121, E-28006
           Madrid, Spain} 
\author{M. Cervi\~no} \affil{Instituto de Astrof\'{\i}sica de Andalucia (CSIC), POB 3004, E-18008 Granada, Spain}
\affil{LAEFF-INTA, POB 50727,
           E-28080 Madrid, Spain} 
\author{Rosa M. Gonz\'alez Delgado} 
\affil{Instituto de Astrof\'{\i}sica de Andalucia (CSIC), POB 3004, E-18008 Granada, Spain}

\begin{abstract}

We present evidence of the co-existence of either an AGN or an
ultraluminous X-ray source (ULX), together with a young super stellar
cluster in the 3 central parsecs of NGC~4303.  The galaxy contains a low
luminosity AGN and hosts a number of starburst regions in a circumnuclear
spiral, as well as in the nucleus itself.  A high spatial resolution
\emph{Chandra} image of this source reveals that the soft X-ray emission
traces the ultraviolet nuclear spiral down to a core, which is unresolved
both in soft and hard X-rays. The astrometry of the X-ray core coincides
with the UV core within the \emph{Chandra} positioning accuracy.  The total
X-ray luminosity of the core, $\sim 1.5\times 10^{39}~$erg\,s$^{-1}$, is
similar to that from some LINERs or from the weakest Seyferts detected so
far.  The soft X-rays in both the core and the extended structure
surrounding it can be well reproduced by evolutionary 
synthesis models (which include the emission expected from single
stars,
the hot diffuse gas, supernova remnants and binary systems), consistent
with the properties of the young stellar clusters identified in the UV.
The hard X-ray tail
detected    in the core spectrum,   however,  most likely requires the
presence of an additional  source.  This additional source could either
be a weak active nucleus black hole or  an ultraluminous X-ray object.
The implications of these results are discussed.
\end{abstract}
\keywords{galaxies: active --- galaxies: individual (NGC 4303) --- galaxies: nuclei --- galaxies: starburst --- X-rays: galaxies}

\section{Introduction}

Seyfert~2  galaxies are a sub-class  of  Active Galactic Nuclei  (AGN)
whose observed properties   suggest that the  supermassive  black hole
that is widely accepted to power AGN is obscured to our line of sight.
Several Seyfert~2 galaxies are   known to host intense  starbursts  in
their  nuclear  regions,  \citep{heckman,rosa,levenson},  which emit a
significant part of their  energy in the UV  and X-ray energy  ranges.
In  most cases  the star formation  episodes  contribute with  a large
fraction of the total UV emission while  the relative contributions of
both  components to X-rays remain  still  to be disentangled.  This is
naturally explained in the unified model frame as due to absorption of
the intrinsic AGN emission by matter in the line  of sight, probably a
dusty molecular torus. The energy required for  the nuclear photons to
show up after going through the obscuring matter mainly depends on its
  effective  HI  column  density.   For columns  of  $10^{22-23}$
cm$^{-2}$, the intrinsic AGN emission is completely absorbed in the UV
and softer X-rays, but can be detected at  medium to hard X-rays.  The
AGN  emission can also  be seen as a  result of reflection.  Seyfert~2
galaxies  and  other type-2  low  luminosity AGN hosting circumnuclear
star-formation knots   provide    therefore an ideal   laboratory   to
investigate    the   starburst-AGN connection,    to  characterize the
intrinsic  X-ray   AGN  emission and   quantify   the contribution  of
starbursts to the whole X-ray emission.  On  one side, the UV and soft
X-ray starburst emission can be properly  analysed thanks to the torus
which  shields (at least  partially) the brightest AGN radiation which
in Seyfert 1 and quasars  is completely hiding possible starbursts. On
the other side,   the AGN itself can still   be characterized via  the
analysis of the hard X-ray emission, provided the torus column density
is lower than $\sim 10^{24}\,cm^{-2}$.

NGC~4303, M61, is a nearby (distance 16.1\,Mpc) widely studied barred
galaxy harbouring a low luminosity AGN. The properties of the optical
emission line ratios classify it as a Seyfert 2/LINER border line AGN
\citep{filippenko,colina-arribas}. A young massive
stellar cluster and a low luminosity active nucleus seem to coexist within
the central 3 pc of the galaxy (1\arcsec = 78 pc), with the starburst
contribution being the dominant one in the UV spectral region.  In
particular, the \emph{STIS-HST} UV spectrum of the core (diameter $\le$ 3 pc)
shows characteristic broad absorption lines produced by massive young star
winds and it can be reproduced by a young massive stellar cluster,
\citep{colina-rosa}, with no need for additional contributions.
\emph{STIS-HST} 
UV images show a circumnuclear spiral-shaped structure of star-forming
regions spread out to a radius of 250-300 pc (i.e. 3\farcs5-4\arcsec).
\emph{HRI-ROSAT} images show that about 80 per cent of the total soft
X-ray   emission   from  NGC~4303, i.e.  $4.7_{-1.4}^{+1.5}   \times
10^{40}\, erg\,s^{-1}$ from  0.1  to 2.4\,keV,  \citep{tschoke}, comes
from the  central ($\sim 25$\arcsec)  region of  the galaxy. The X-ray
contours reveal that this central source is possibly extended, but the
detailed (and smaller scale) circumnuclear spiral discovered in the UV
cannot be resolved with HRI.   The question whether this central X-ray
emission is  from pure star-forming massive  clusters or a composition
of these star clusters plus a low luminosity AGN is not solved, although
the second explanation is suggested by \citet{tschoke}.

We have obtained \emph{Chandra} images of NGC~4303, aiming to
resolve the X-ray emission from the core and from the circumnuclear
star-forming spiral. In this paper, we discuss the spectral properties
of both 
regions and model the X-ray emission associated to the known star-forming
knots with evolutionary synthesis models. In Sect. 2 we present the \emph{
Chandra} data and their analysis. In Sect. 3 we discuss the origin of the
X-ray emission and its relation to the star-forming episodes. In Sect. 4 we
summarize the main conclusions of our work. 

\section{Data analysis and results}

\subsection{Observations}
In this  paper we present  the  \emph{Chandra} Advanced CCD  Imaging
Spectrometer (ACIS) observation  of  NGC~4303. ACIS covers  the energy
range between 0.2-10 keV. Each of the 6  active CCDs of the instrument
observed a 8\farcm3$\times$8\farcm3 field of view. NGC~4303 was chosen
to  be placed in  the S3 back illuminated ACIS  chip. Around 90$\%$ of
the encircled energy lies within  4 pixel in diameter (2\arcsec). The
observation was performed  on August 7$^{th}$ 2001 with a total duration
of the standard good time intervals of 28380\,s. Data  processing and
calibration was performed with the \emph{Chandra} software CIAO 2.1. 
The analysis has been done using the the level 2 event lists as 
generated by the standard processing system, on August 20, 2001. The only
further correction that was needed and applied was the use of improved gain
maps made available after the processing date. 

\subsection{Morphology of the X-ray emission}

\subsubsection{Off-nuclear emission}\label{off-nuclear}

\begin{figure}
\plotone{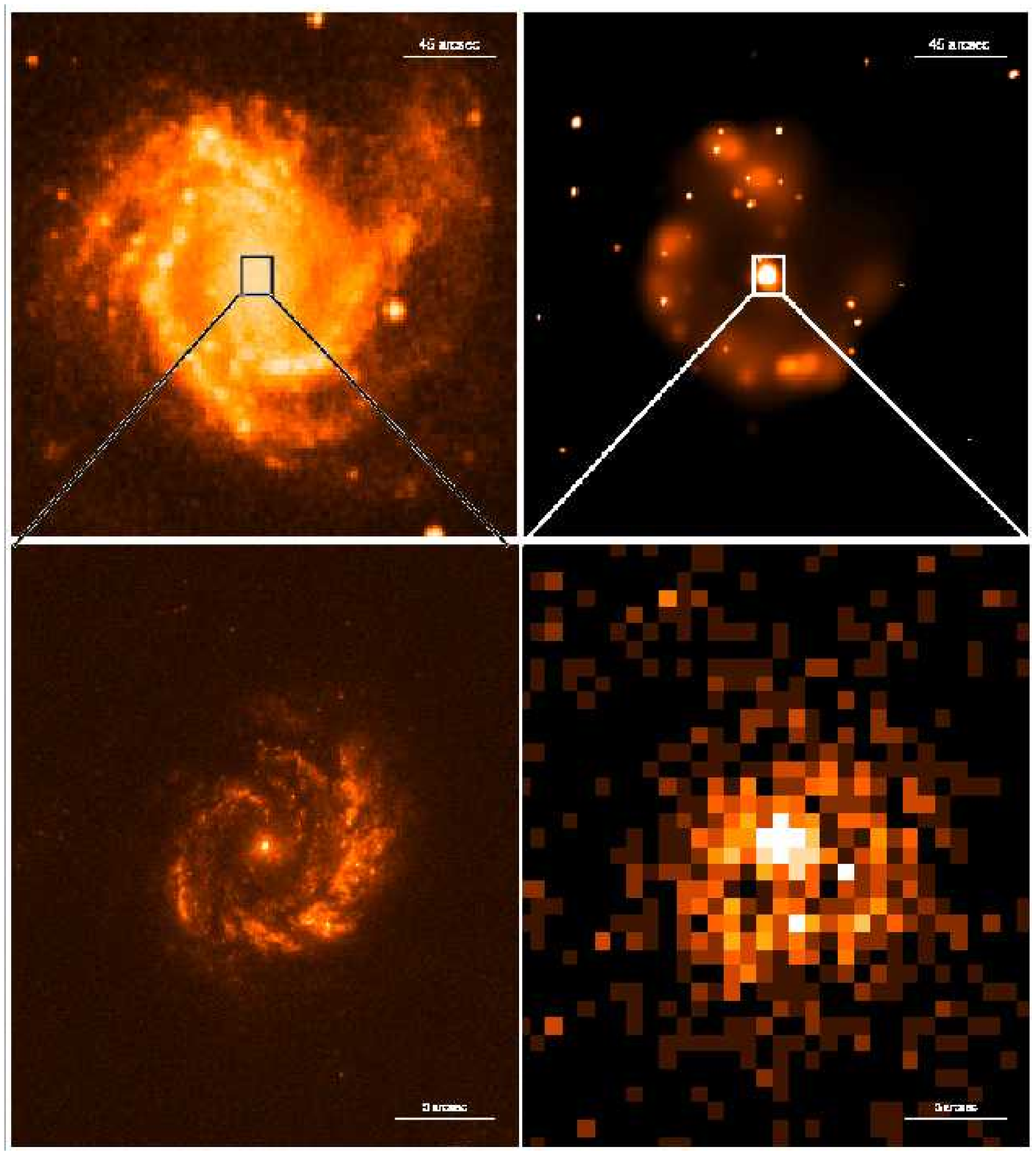}
\caption{
{\bf Top Images:} Comparison between DSS (on the left) image and
smoothed \emph{Chandra} image (on the right) of the  same region. The spiral structure
in the  DSS   coincides  with the  structure  observed in  the  
\emph{Chandra} image.  {\bf Low images:} \emph{HST} UV and \emph{Chandra}
images of the nuclear region are also shown, (on left and right, respectively). North is up and East is  left.\label{chandra_dss}}
\end{figure}

The high spatial resolution of the ACIS image allows us to distinguish
many features.   The image   shows structure  in the  intense  nuclear
region of  NGC~4303,  surrounded by a  dim halo  with  an extension of
about  2\arcmin , mapping  the most luminous arm   of the galaxy.  The
5\arcmin$\times$5\arcmin\/  \emph{Chandra}  image  of  NGC~4303 can be
seen  in Figure~\ref{chandra_dss}, compared   with a \emph{Digital Sky
Survey} (DSS) image of the same region (top panel).  The dim structure
in the \emph{Chandra} image is mapping the spiral arms seen in the DSS
image.      As  discussed     bellow,   in       Sections~\ref{fluxes}
and~\ref{discussion},  the      total      diffuse    emission      is
$1.3\pm0.4\times10^{40}$\,erg\,s$^{-1}$.

\begin{figure}
\plotone{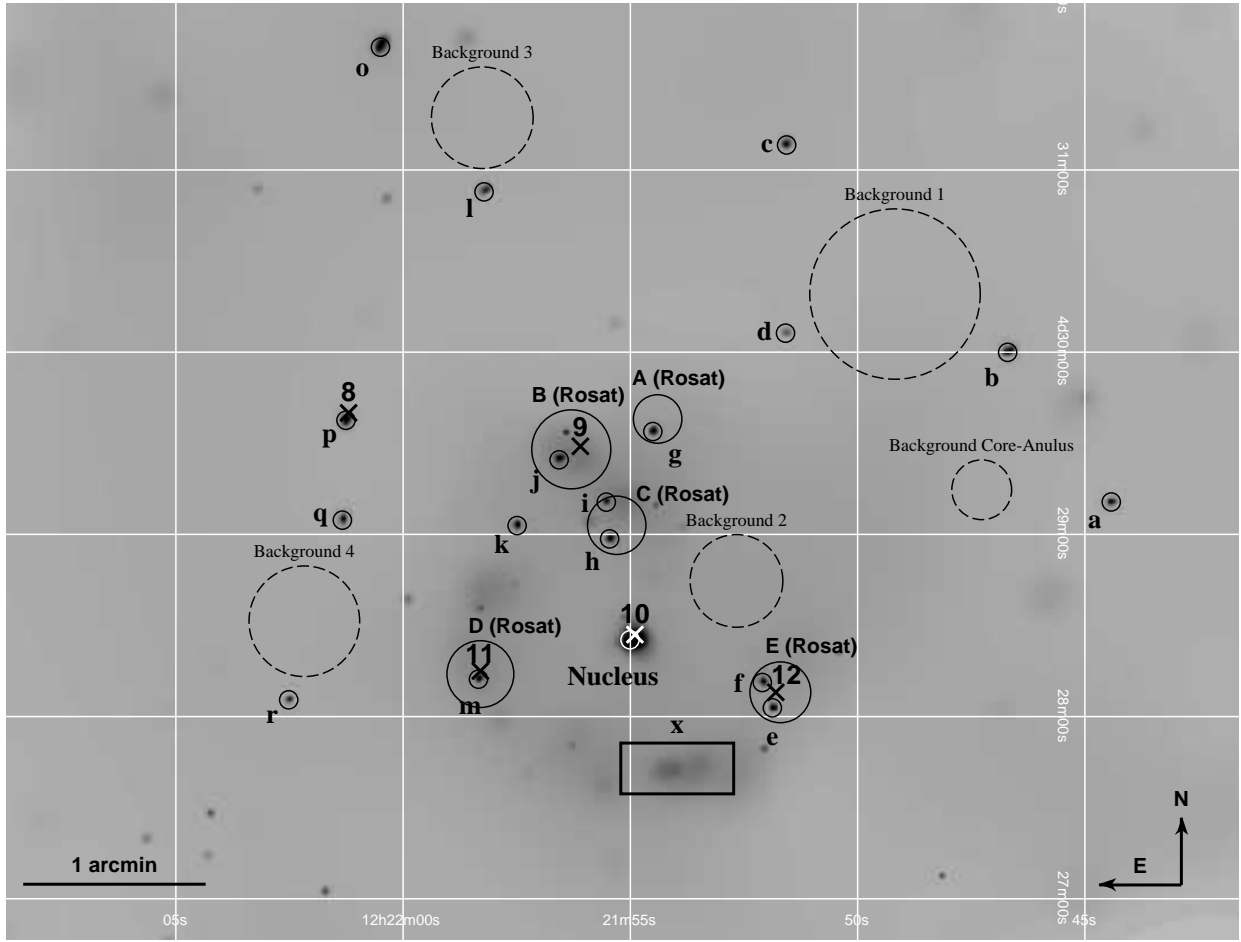}
\caption{ Location of the different sources detected by \emph{Chandra}
(small circles) and
\emph{ROSAT} (numbers and big circles) placed in the \emph{Chandra}
field. The background regions used for the analysis of the different
point-like sources, the core and the annulus are also shown. North is
up and East is left.\label{sources}}
\end{figure}

In addition to the diffuse emission, some  point-like sources are seen
in  X-rays, probably  associated   with star-forming  knots  or binary
systems.  These  sources are shown   in in Fig.~\ref{sources} together
with their \emph{ROSAT} identifications, following the nomenclature of
\citet{tschoke},  when  applicable. The  location, net  counts and net
count    rate  of    each      detected    source  is     shown     in
Table~\ref{table_sources}.  The sources have  been detected using  the
CIAO  2.1 \emph{celldetect} detection tool.  The counts of each source
have been   calculated integrating over a   circle of 4  {\it Chandra}
pixels.   Four  different background regions   have been considered in
order   to obtain the net  counts,   see Table~\ref{table_sources} and
Figure~\ref{sources}. The  errors considered for the count calculation
are       poissonian,   i.e.     using   the    Gehrels  approximation
$\Delta\textrm{counts}=1+\sqrt{\textrm{counts}+0.75}$, \cite{gehrels}.
The errors for the net counts  were calculated assuming that the count
error   for    the   source    and    for     the  background      are
independent. Unfortunately, the signal to noise is  not high enough to
allow individual spectral  analysis  for all  the sources.  To compute
their luminosities  in the 2--10\,keV energy  range, we  have used the
count  rate   estimator tool,  {\it PIMMS~3.2d}.   The point like
sources detected in the \emph{Chandra} image are located in the spiral
arms of NGC~4303. The sources coincide with the HII regions studied by
\citet{martin} or are located near them. These regions seem to be
photoionized  by      OB      stars   related      to       starbursts
episodes,\citet{martin}.  Since  these   sources  are   located within
star-forming regions, they are  likely to be  high mass X-ray binaries
(HMXB). According to \citet{persic}, HMXB spectra  can be modeled by a
power law with  index 1.2$\pm$0.2 in  the 2--10\,keV range.  This
value is also in agreement with  the indexes found  for HMXBs in Large
Magellanic Cloud, ($\Gamma=1.0-1.4$, \citet{haberl})  and in the Small
Magellanic Cloud,($\Gamma=1.0-1.7$, \citet{sasaki}), observed with
\emph{XMM-Newton}.  However, the luminosities   for these sources  are
one or  two orders of magnitude lower  than the luminosites calculated
for  the NGC~4303 point-like sources.  We have used this model, power
law with $\Gamma=1.2\pm0.2$, to derive the luminosities of the sources
in this   energy band.  The  computed  luminosities  were corrected by
line-of-sight  absorption  using    the Galactic  column   density  of
$1.67\times 10^{20} {\rm cm}^{-2}$ \citep{schlegel}  and are shown  in
Table~\ref{table_sources}.  The  errors in the luminosities  have been
estimated using the error in the power law  index.  The previous power
law model does  not apply  in  the soft, 0.23--2\,keV, band.   We have
therefore computed an energy  conversion factor, ECF, by performing an
spectral analysis of  the brightest point-like source  detected within
the  central   2\,\arcmin~region.  This    is    source {\bf  m}    in
Fig.~\ref{sources},  for which we   have  extracted the  spectrum  and
fitted    an     absorbed,  n$_H$=$5^{+7}_{-2}\times10^{20}\,cm^{-2}$,
Raymond-Smith thermal    model, kT=$2.6^{+1.9}_{-0.9}\,keV$ with fixed
Galactic absorption.  We have  obtained a value for the ECF$_{0.23-2}$
of  $5^{+1}_{-4}\times10^{-12}$~erg\,cm$^{-2}$\,s$^{-1}$counts$^{-1}$,
where the error  has been estimated using  the uncertainty in the flux
of the source {\bf m} obtained from the fit.

The  luminosities  found in  both  energy    bands are $\gtrsim
10^{38}\,ergs^{-1}$ for  the majority of the  sources, i.e.  above the
Eddigton         limit    for       a      typical      neutron  star,
$L_{Edd}=1.5\times10^{38}\frac{M}{M_\odot}\,ergs^{-1}$     and   these
sources could be  associated to intermediate  X-ray objects, (IXOs) or
ultraluminous X-ray sources, (ULXs).  Recent works find that this kind
of sources can be explained by a power law with index within the range
1.6-2.9\footnote{\emph{Chandra}   data     of  two  of  the  brightest
point-like sources in nearby  galaxies, NGC~5204~X-1 and NGC~4559~X-1,
shows indexes of $\Gamma=1.9-2.2$ and $\Gamma=2.4-2.9$, respectively,
\citep{roberts}. \citet{strickland} found a value of the power law
index of $\Gamma=1.6-2.0$ for an ULX  in NGC~3628, observed with \emph
{Chandra.} \citet{foschini} studied with \emph{XMM-Newton} a sample of
8 ULXs, finding that  5  of them could be   model by a power law  with
$\Gamma=1.9-2.3$. The  fits have been preformed with \emph{XMM-Newton}
and \emph{Chandra}   data, so they are  valid  in both   energy bands:
0.23-2~keV and 2-10~keV.}, also in  agreement with the values for soft
and  very  hard   states   of black hole   X-ray   binaries,  (BHXBs),
$\Gamma\sim2.5$.  We have  therefore computed also the luminosities in
the two energy bands considered using the power law index $\Gamma=1.6$
and   $\Gamma=2.9$.    The    resulting     values  are    shown    in
Table~\ref{table_sources} with the  luminosities marked with L$^2$ and
L$^3$.   Unfortunately,  the quality  of the   data  does not allow to
perform  an individual spectral analysis of  each source, necessary to
determine the nature of the off-nuclear point-like sources.

In addition to the point sources, we have also measured the counts and
estimated the fluxes and luminosities of an extended region in the
circumnuclear ring. This is the region labeled with an {\bf x} in
Fig.~\ref{sources}, centred at RA=12$^{\rm h} 21^{\rm m} 54\fs0 $ and 
DEC=+4$\degr 27\arcmin 42\farcs9 $. The value of the net counts
detected in
this region in the 0.23--2\,keV band is 291$\pm$19, suggesting, using
the
above ECF, a soft X-ray luminosity of $16.1^{+ 4.3}_{-14.0}\times10^{38}$\,erg\,s$^{-1}$. 

\subsubsection{Spatial analysis of the nuclear region}

\begin{figure}
\includegraphics[angle=-90,scale=1.5]{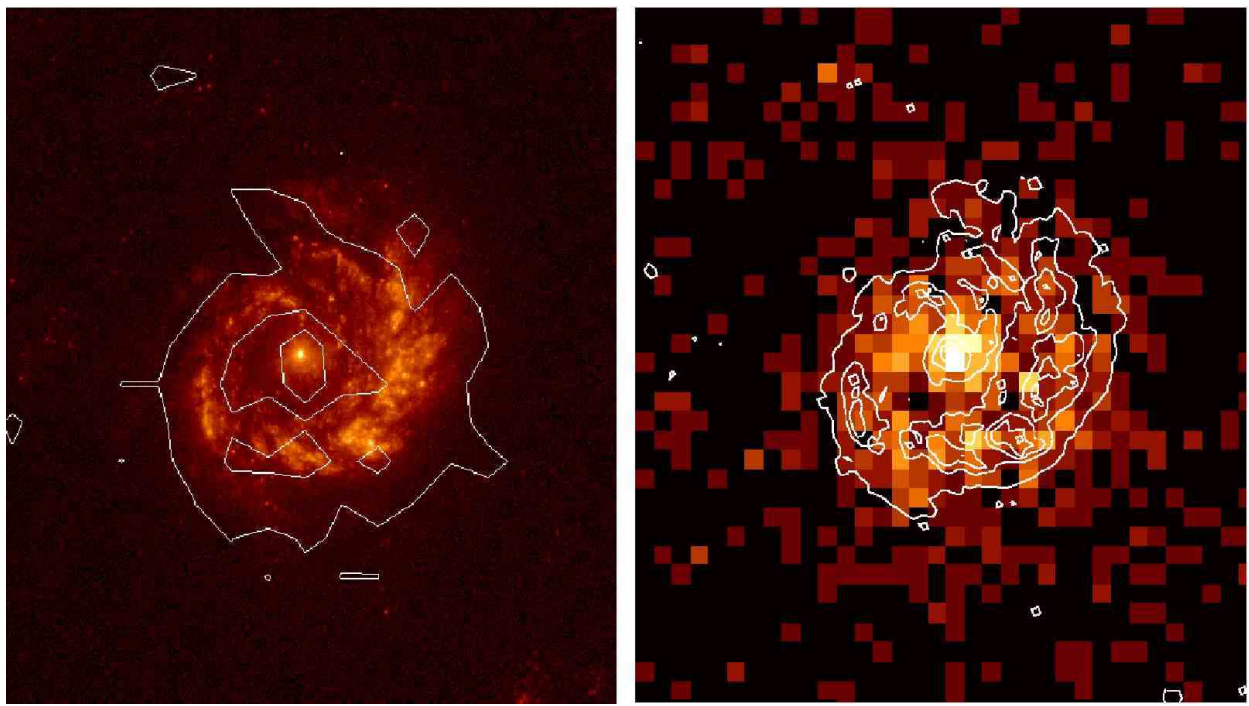}
\caption{ \emph{Chandra} image with \emph{HST}
contours  superimposed (on the  left) and \emph{STIS-HST} image with \emph{Chandra} contours
superimposed  (on the right).  The   NGC 4303 central  region
shows a small   spiral  structure. The \emph{Chandra}  and  \emph{HST}
structures coincide.  North  is up and East  is left. The size of each
image is 17\arcsec$\times$ 18\arcsec.\label{hst_chandra}}
\end{figure}

The central region of the galaxy shows a well defined X-ray structure when
looked closer, (see Figure~\ref{hst_chandra}, on the right).  The X-ray
image has been compared with a high resolution \emph{STIS-HST} UV image
\citep{colina-rosa}.  The UV image shows a luminous nucleus surrounded by a
nuclear spiral structure of star forming regions.  The \emph{Chandra}
contour map superimposed on the \emph{STIS-HST} image (or, similarly, the
\emph{STIS} contours superimposed on the \emph{Chandra} image) shows that
there is a good match between UV and X-ray emission regions (see
Figure~\ref{hst_chandra}, on the left).
Moreover, the positions of the unresolved core in both ranges agree better
than the \emph{Chandra} astrometric accuracy of 0\farcs6 (90\% confidence):
whereas the core location in the \emph{HST} image is 
RA (J2000)=12$^h$ 21$^m$ 54\fs96 and
Dec(J2000)=+04\degr28\arcmin25\farcs52 \citep{colina-rosa}, in the X-ray
image it is located at RA (J2000)=12$^h$21$^m$ 54\fs95 and
Dec(J2000)=+04\degr28\arcmin25\farcs7.  This means that the UV nucleus of
NGC 4303 is coincident with the soft and hard X-ray nucleus within the
astrometric accuracy of \emph{Chandra} the  \emph{STIS-HST} accuracy being much
better than this.

\begin{figure}
\epsscale{0.7}
\plotone{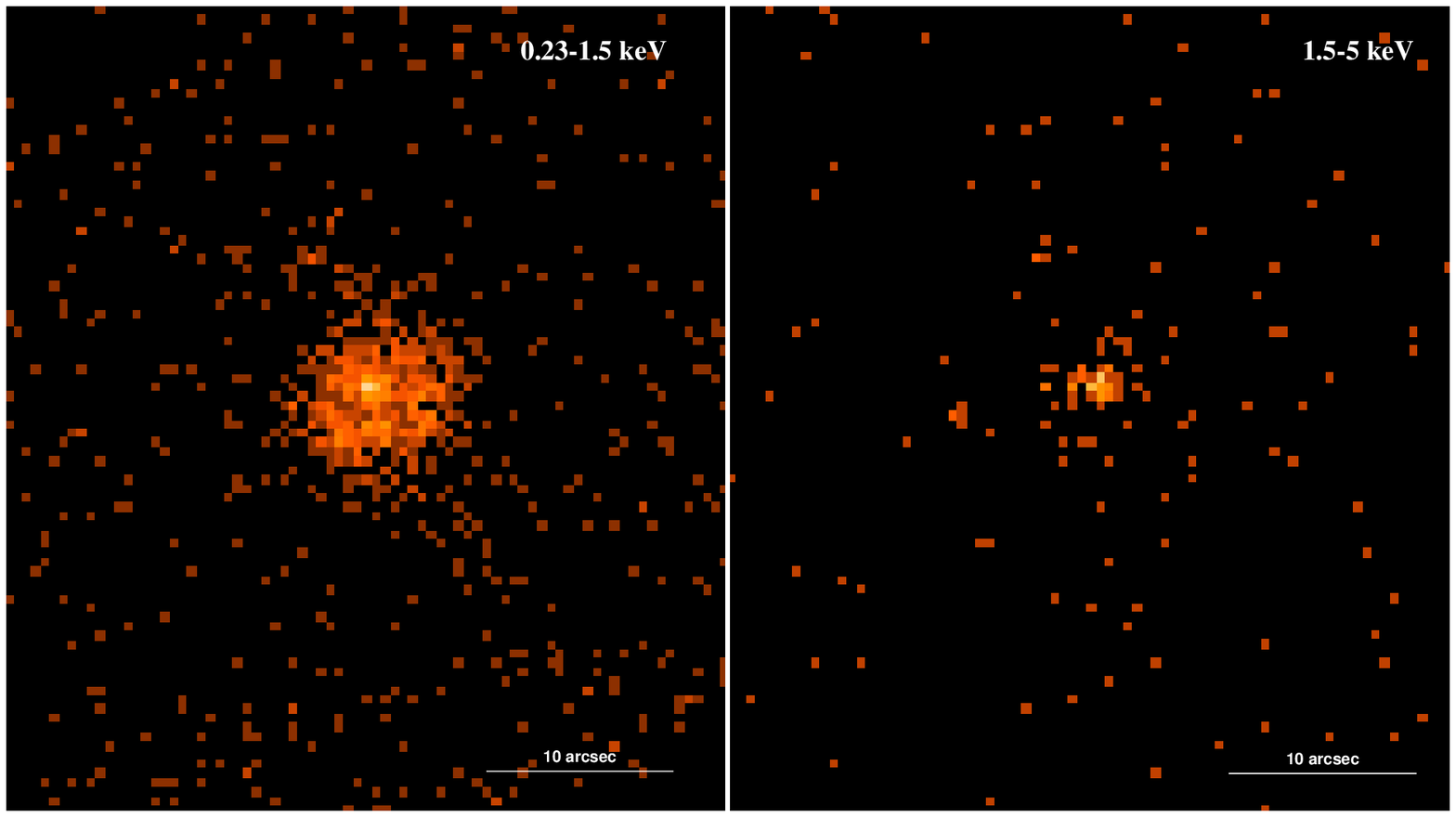}
\epsscale{0.7}
\plottwo{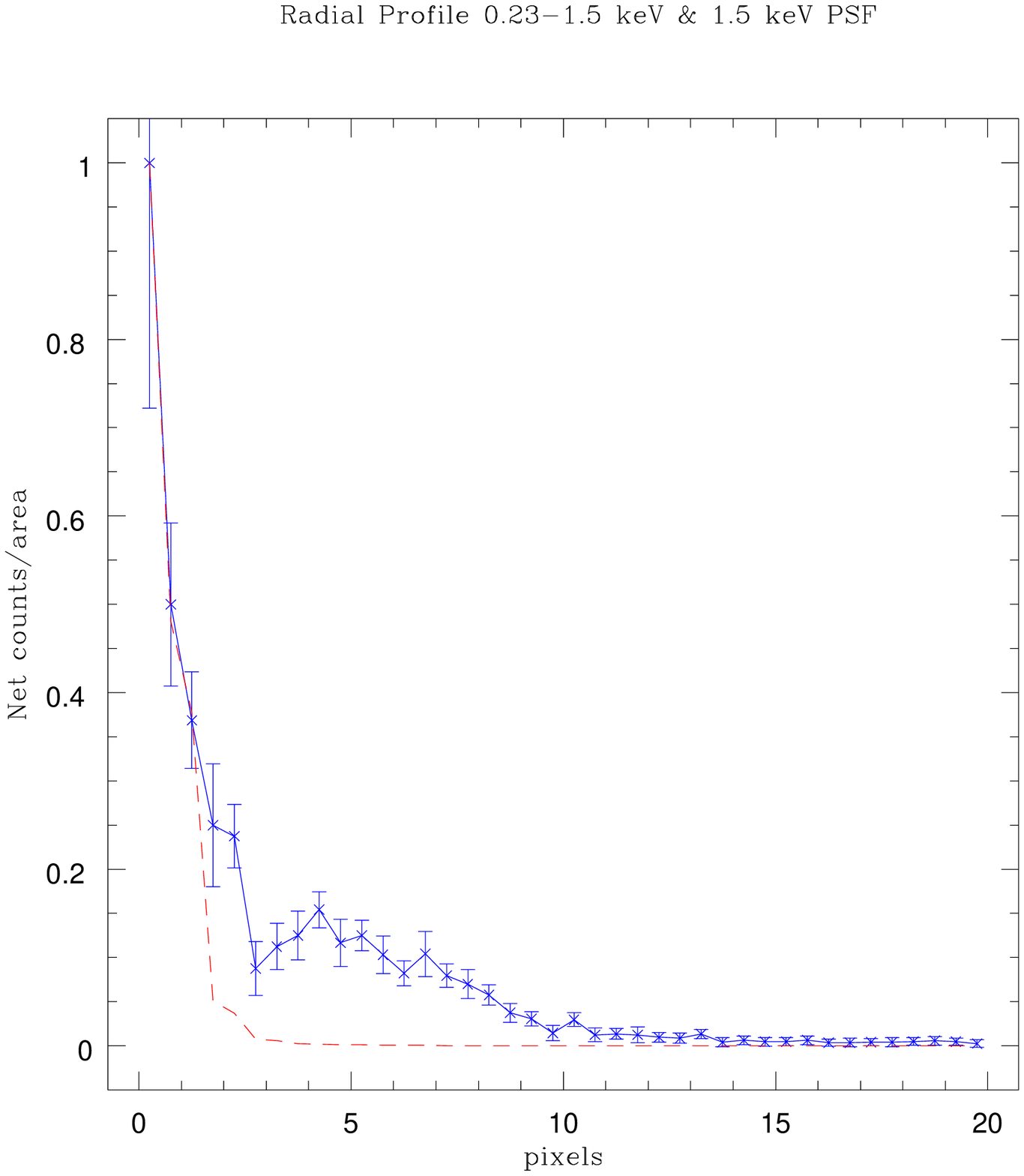}{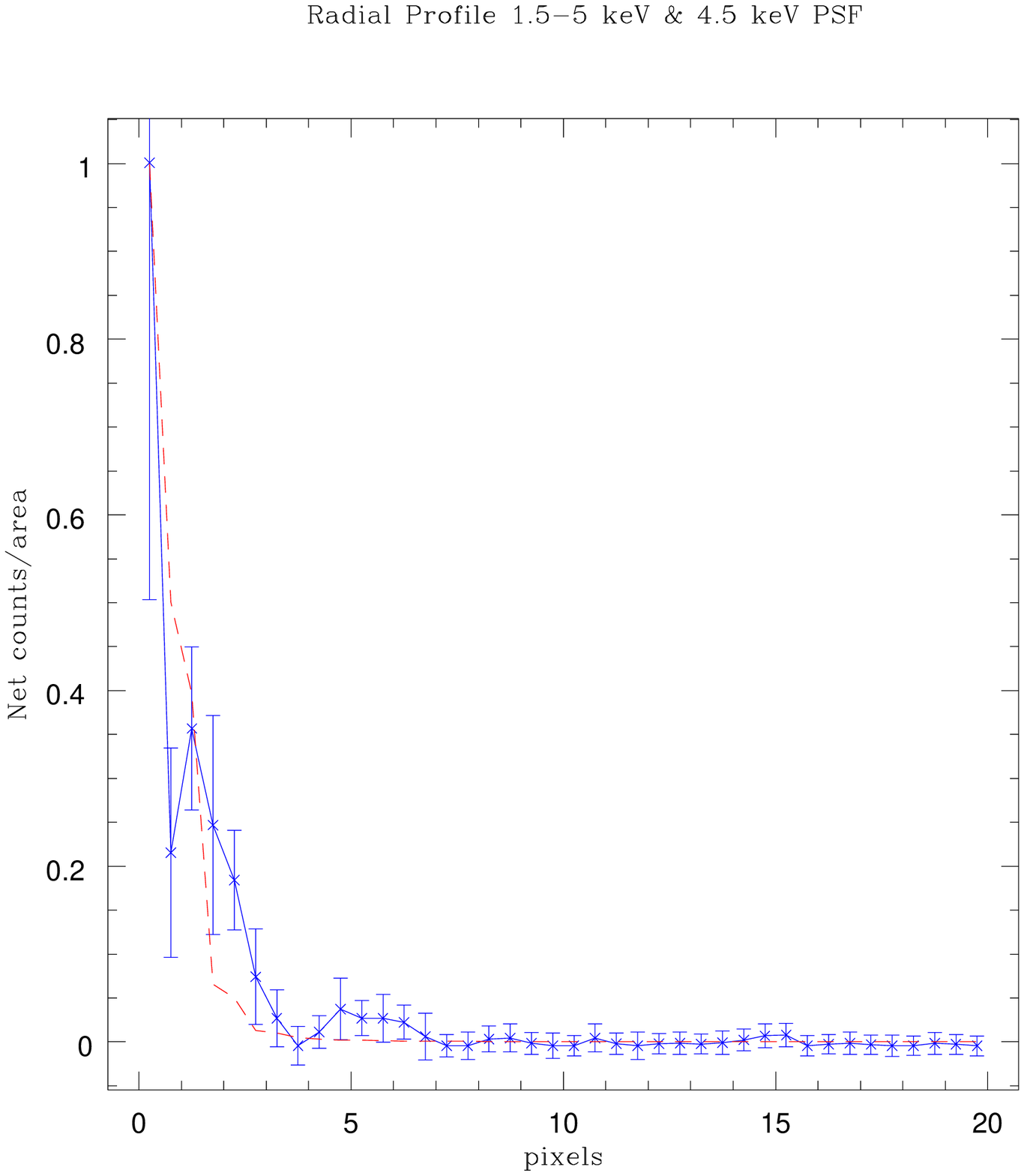}
\caption{ 
 On  the  top, \emph{Chandra}   images of   NGC~4303 in two  different
 bands. On  the  left, 0.23-1.5 keV; on   the  right, 1.5-5  keV.  The
 dimension of both images is 40\arcsec $\times$ 40\arcsec. North is up
 and East is left. Underneath, the radial profile of the central 20 px
 ($\sim$ 10\arcsec) of NGC~4303 for the  0.23-1.5 and 1.5-5 keV bands,
 compared    with  the    expected  PSF   (dashed   line).    1  px  =
 0\farcs492.\label{bands_rprofiles}}
\end{figure}

As a first step for the  spatial analysis of the nuclear region,
broad band soft, 0.23-1.5 keV, and  hard, 1.5-5 keV, X-ray images have
been compared.   The energy ranges have  been  chosen according to the
observed   spectrum  and   \emph{Chandra}  effective  area, to  better
separate the  soft and  the hard emission    while maintaining a  high
enough effective area and measured counts in the hardest band to allow
computation of hardness  ratios.  The radial  profiles have  also been
extracted and  compared with the point  spread function.   The results
are shown in  Figure~\ref{bands_rprofiles}.  In  the hard X-ray  image
only the most central region  is present and no significant difference
is found between the central NGC~4303 profile  and the PSF; this means
that  the hard  X-ray source is   unresolved.  Meanwhile, the  spatial
structure of the NGC~4303 central region  clearly shows up in the soft
X-rays. The radial distribution of  the low energy emission is clearly
different from what is expected from a point-like source. The position
of the  peak intensity coincides with  the unresolved hard  X-ray one,
but there is a secondary peak close to 2\arcsec\ from it and extending
at least  to $\sim$ 8\arcsec\ away.  Based  on  these results, we have
defined two regions: the \emph{core} and the
\emph{annular region} as the 1\farcs48 (3 \emph{Chandra} pixels) radius
circle and the annulus between 2\farcs46 and 7\farcs87 (i.e.  5 and 16
\emph{Chandra} pixels, in radius), respectively.  The inner radius of the
\emph{annular region} has been set to 5 pixels to avoid overlapping
pixels between the two zones. The outer radius is big enough to consider
all the emission 
originating in the annulus, but excluding the external halo.

\subsection{Spectral analysis}

\subsubsection{Broad band properties}

In order to quantify the spectral differences between the core and the
annulus,       we     have computed   the      hardness     ratios as:
$HR=(hard-soft)/(soft+hard)$, where \emph{soft} and \emph{hard} are
the  net   count rates  integrated   in  the  0.23--1.5\,keV  and
1.5--5.\,keV  energy bands, respectively. We have measured \emph{soft}
values of $151\pm12$ and $398\pm20$ counts, for the core and annulus,
respectively and, similarly, \emph{hard} values of $53\pm7$ and $31\pm6$
counts. With these values, we obtain $HR_{core}=-0.48\pm0.06$
and  $HR_{annulus}=-0.86\pm0.03$. 
This means that about 25\% of the X-ray counts from the core have energies
above 1.5\,keV, in   contrast to  7\%  of   the  detected counts   from the
annulus.  The errors on the $HR$ have been calculated assuming the count
rates of each band  to be independent measurements.  The difference
between the 
two values of $HR$ equals to 5.7$\sigma$, with $\sigma$ its error.

\subsubsection{Core region}\label{score}

The total net counts and net count rate in  the \emph{core} region are
188  counts     and    $(7.4\pm   0.5)\times     10^{-3}\,    {\textrm
counts\,s}^{-1}$, respectively.  We  have  extracted the  spectrum and
grouped it based on a requirement  for 25 counts  per channel to allow
use of $\chi^2$ statistics. The  analysis has been performed with  the
\emph{Xspec}  software package, using as  background the spectrum of a
3\,pix  radius circular  region  centered about  2\,\arcmin\,NW of the
nucleus, see Fig.~\ref{sources}.

The  first  trial has been  to  fit single component models,  either a
power  law or a thermal Raymond-Smith  model, absorbed by the Galactic
column   density, N$_{\rm H}=1.67\times 10^{20}\,{\rm cm}^{-2}$
\citep{schlegel},  and with only the normalization  and the relevant parameter  (photon index
of the  power law, $\Gamma$, or temperature  of the thermal component,
T), left free, as the  abundance has been frozen  to the solar  value.
None of the fits were statistically acceptable with $\chi ^2_{\nu} \ge
2.4$ (6 degrees of freedom, dof), see Table~\ref{table_core}.  In
both cases, the  residuals of the fit  are larger  in the soft
energy  range, $<$~2~keV. Two-component  models
have therefore being    tested,  though keeping the number    of  free
parameters  to  the  very minimum,   given the  small  number  of bins
available.

\begin{figure}
\includegraphics[angle=-90,scale=0.7]{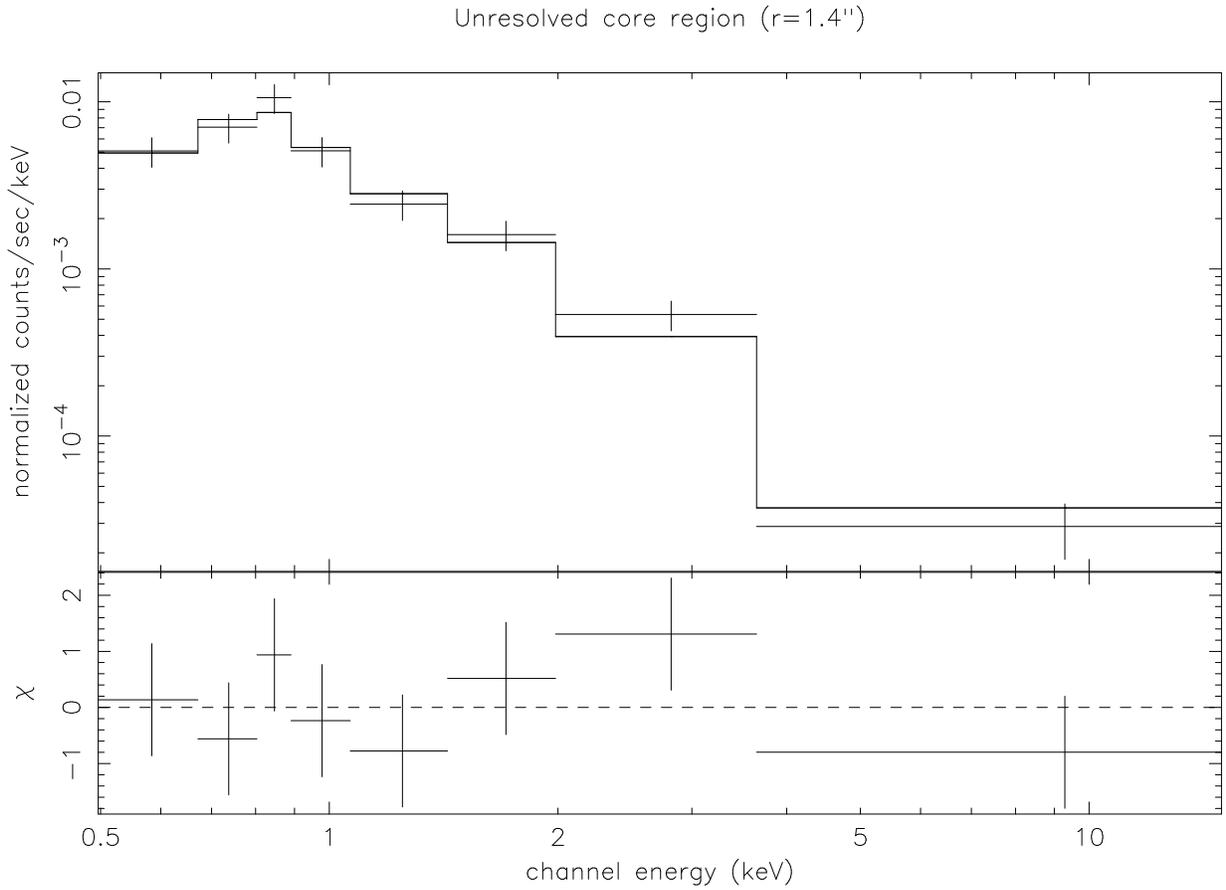}
\caption{The \emph{Chandra} spectrum of the unresolved core, the best
fit model and  the $\chi$ residuals. The model consists of a power 
law and a thermal Raymond-Smith component, both absorbed
by the Galactic column density.  See Table~\ref{table_core} for the 
model details. \label{core}}
\end{figure}

The  best fitting model found for   this region, $\chi^2_{\nu}$=1.1 (4
dof), includes two components: the emission  of a hot and diffuse gas,
Raymond-Smith   emission,  and a  power    law,   with only  the   two
normalizations, T and  $\Gamma$ let free. Photoelectric absorption due
to our  galaxy has  been added  with n$_{\rm  H}$  fixed to the  above
value.     We have tested the  possibility   of an additional internal
absorption for the  thermal component,  obtaining  a very low   value,
n$_H\sim10^{11-12}\,{\rm cm}^{-2}$; therefore, it can be neglected and
it is not considered  any further.  The values of  $\chi^2$ and dof of
the  single component  models and  the best   fitting model have  been
compared using the F-test.  The value  of F obtained is 4.54, assuring
the  improvement obtained with the  two component fit is statistically
significant with  a probability  of  90\%.  Figure~\ref{core} shows  a
plot of the
\emph{Chandra} spectrum, the best fit model and the $\chi$ residuals.

An alternative model, with a thermal Raymond-Smith plus a bremsstrahlung
component, as generally used for AGN accreting with low radiative
efficiency, is formally compatible with the above fit.  The $\chi^2_{\nu}$
is 0.99 for 4 dof, the Raymond-Smith has the same temperature as in the
previous fit: kT=$0.6^{+0.2}_{-0.3}\, keV$ and accounts for the soft
X-rays, while the hard X-rays are explained with a bremsstrahlung
temperature of kT=$8^{+60}_{-5}\,keV$. The main problem of this model
is the very large uncertainty on the latter parameter.  Another fit was
tried with a thermal component and a power law whose photon index was fixed
to 1.9, the mean value found in radio quiet type 1 AGN, and allowing
variable power law absorption.  The best fit parameters were a higher column
density, $8\times 10^{20}\,cm^{-2}$, as expected for type 2 AGNs, and a
thermal component temperature of 0.6\,keV, similar to the result in the
previous fits. However, it failed to give an statistically acceptable
result, $\chi^2_{\nu}=1.34$ for 4 dof and therefore it cannot be considered
as an adequate model.

Last, we have also tried to fit models that give good results with spectra
of ultraluminous X-ray sources, ULX. These models consider an optically
thick accretion disk consisting on multiple blackbody components, the
multicolor disks (MCD), and can often explain either alone or in
combination with a power law, the ULX emission \citep{makishima}. Neither a
single MCD fit, nor MCD plus a power law fit have given statistically
acceptable results, nor even if some of the parameters have been fixed to
typical values found for other ULX (e.g. temperature at the inner disk
radius, $\textrm{T}_{in}$, in the range of 1--1.8\,keV \citep{mitsuda} or
power law index 1.5--2.). The best fit obtained with MCD has a
$\chi^2_{\nu}\sim 4$ for 6 dof, and it improves adding a power law
component, but the value of $\chi^2_{\nu}$ is still unacceptable,
$\ge3$. The values of the parameters of each model and the goodness of
the fits are shown in Table~\ref{table_core}.

\subsubsection{Annular Region}

The total net counts and net count rate in  the annulus are 477 counts
and    $(1.6  \pm   0.1)\times    10^{-2}\,{\textrm  counts\,s}^{-1}$,
respectively. The spectrum has been extracted  and binned to a minimum
of 35 counts per bin. The  background spectrum has been extracted from
a region centred  in the same position as  the background used for the
core spectrum, but with  different geometry, i.e. a  ring of inner and
outer radius of 5 and 16 pixels, respectively, Fig.~\ref{sources}.

\begin{figure}
\includegraphics[angle=-90,scale=0.7]{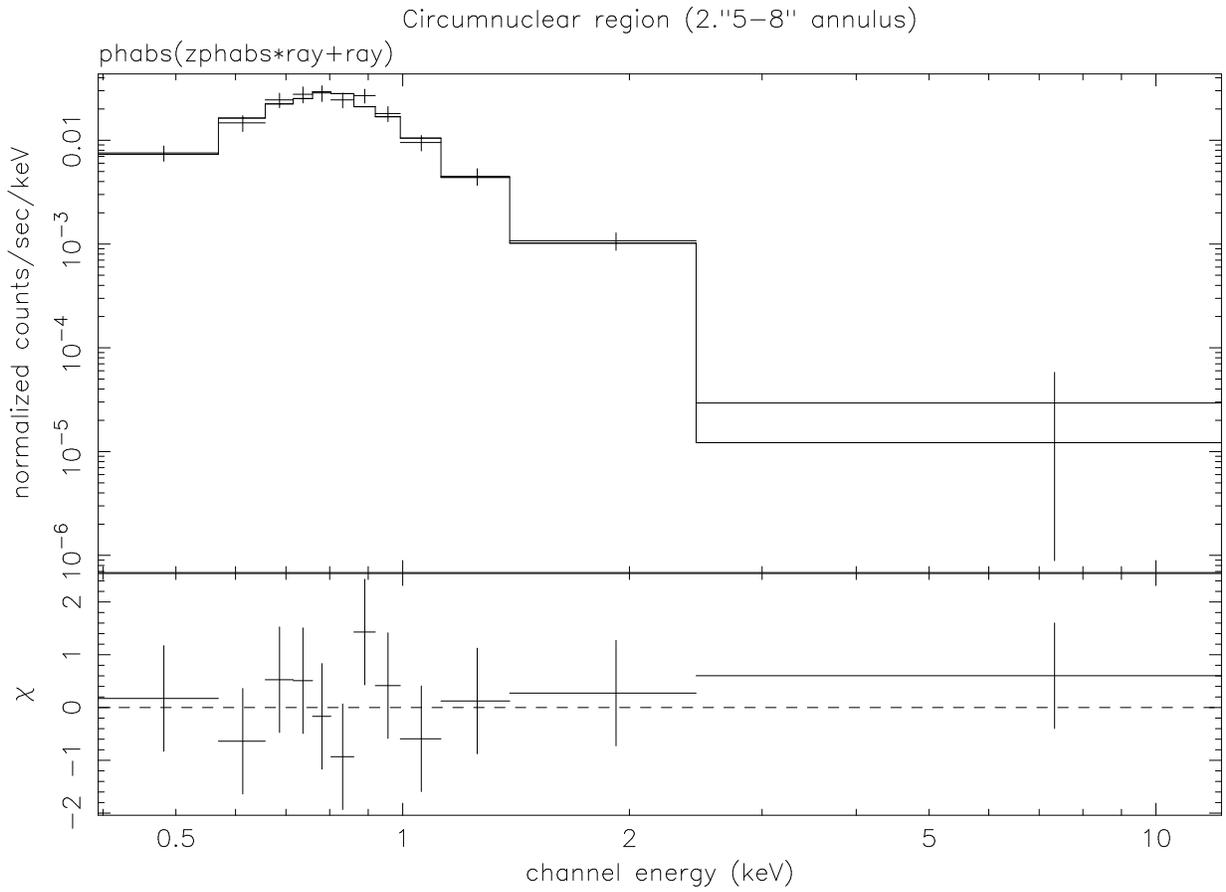}
\caption{The \emph{Chandra} spectrum of the annular region, the best
fit model and the $\chi$ residuals. The model consists of two thermal 
Raymond-Smith components with different absorption, see 
Table~\ref{table_annulus} for the model details. \label{annulus}}
\end{figure}

Similarly to the core, single model  fits give unacceptable results.
A fit with  a combination  of a  thermal  model plus a power  law
gives a $\chi_\nu^2=1.12$ for 7 dof. The best 2-component fit uses two
thermal  Raymond-Smith     models,    $T_I=0.8^{+0.1}_{-0.2}$   and
$T_{II}=0.31^{+0.06}_{-0.05}$, both  absorbed   by  the  Galactic  column
density,  but  the    first  one requiring    an additional  internal
absorption,      n$_H=4.9^{+3.0}_{-2.5}\times10^{21}\,cm^{-2}$.     The
$\chi_\nu^2$ was 0.70 for 7 degrees of freedom. The values of all
model parameters are shown in Table~\ref{table_annulus}. The data, the
best    fit model   and  the      $\chi$  residuals   are plotted   in
Figure~\ref{annulus}.

\vspace{3cm}
\subsection{Fluxes and luminosities}\label{fluxes}

We have computed the fluxes and luminosities of the core and the annular
regions. 
Table~\ref{table_lum} shows the luminosities and fluxes in two different
bands, 0.23-2 keV and 2-10 keV, for each region, corrected from 
absorption, as well as the contributions of each component to
the total emission.  The bands have been chosen to allow comparison with
\emph{ASCA} and \emph{ROSAT} data, either for NGC~4303 or for other
objects of some different classifications. 
The analysis shows
that the emission from the annulus dominates the total emission of the
8\arcsec~internal region in the soft range, contributing with $\sim$ 80 $\%$ of the
total flux in this range, and it is thermal in origin.  On the contrary, in
the 2-10 keV band, the emission is strongly dominated by the core which
contributes with $\sim 90\,\%$ of the hard flux, and follows a power law
distribution.  This result is supported by the images in
Figure~\ref{bands_rprofiles}.

In order to compare our results with previous \emph{ROSAT} analysis, a
spectrum of a region of the same extraction radius as the \emph{PSPC-ROSAT}
spectrum, r=100\arcsec, has been analyzed and fitted in the 0.23--2 keV energy
range. The goodness of the fit obtained was $\chi^2_{\nu}=1.82$ for 50
degrees of freedom. The best model combines a power law with $\Gamma=1.2$
and 2 Raymond-Smith thermal components with temperatures $0.25 \pm 0.02$
and $0.79 \pm 0.06$ keV, the second one absorbed by a column density of $(9
\pm 1)\times10^{20}\, cm^{-2}$.  The flux measured by \emph{Chandra} in
the  0.23-2  keV  is shown in   Table~\ref{table_lum}.  The unabsorbed
luminosity   measured by   the  \emph{ROSAT}~PSPC,  $4\pm 1.5  \times
10^{40}\, {\textrm erg\,s}^{-1}$ in the 0.07-2.4 keV band
\citep{tschoke}, is consistent within the statistical uncertainties with
the \emph{Chandra} 0.23-2 keV measurement across the same aperture
($1.8^{+0.4}_{-0.3}\times10^{40}\,erg\,s^{-1}$).

\section{Discussion}\label{discussion}

As  presented above, NGC~4303   \emph{Chandra}  images reveal  that  a
significant part of its resolved X-ray emission comes from the central
($r \le 8$\arcsec) region.  There is, in addition, a diffuse component
extending to   about a radius  of  $r \sim  100$\arcsec\  plus several
point-like sources.  The luminosity  of the 9 sources  detected within
the  100\arcsec (see Fig.~\ref{sources})   adds up to $\sim$~1.2 times
the luminosity of the central, core plus annulus, region in the in the
0.23--2\,keV band and to $\sim$~1.6 times of the central 2-10 keV band
luminosity.   The total    extended,  diffuse,  soft  X-ray  emission,
($1.2\pm0.4\times10^{40}$\,erg\,s$^{-1}$),     is       a       factor
$2.9^{+1.4}_{-2.5}$ larger than the central soft X-ray luminosity.

\citet{mas-hesse} have analyzed a sample of 111 galaxies, including
QSO, Seyfert 1 and 2 and star forming galaxies, studying the ratio of the
soft X-ray, L$_{0.5-4.5\,keV}$, and Far Infrared luminosities, L$_{FIR}$.
This parameter spans over almost 3 orders of magnitude for 
the different types of galaxies. In order to compare with the average
ratios given by these authors, we have computed it
for the whole NGC~4303 galaxy: the soft X~ray flux has been
integrated from the \emph{Chandra} spectrum for an 100\arcsec\ region,
while the Far Infrared flux F$_{FIR}$ corresponds to the 
IRAS flux at 60~$\mu m$ (as assumed by \citep{mas-hesse}), obtaining 
F$_{0.5-4.5\,keV}=6.5\times10^{-13}\,{\rm erg\,s^{-1}\,cm^{-2}}$ and 
F$_{60\mu m}=23.6\pm1.7$ Jy \citep{iras}. These values yield a ratio
$\log{(\frac{L_{0.5-4.5\,keV}}{L_{60\mu m}})}=-3.3^{+0.1}_{-0.2}$, 
in very good agreement with the average value for galaxies dominated by
star formation, $\log{(\frac{L_{0.5-4.5\,keV}}{L_{60\mu m}})}_{SFG}=-3.33$. 
NGC~4303 as a whole looks therefore like a normal spiral galaxy. 

The X-ray emission of the central region shows a clear structure that
matches very closely the nuclear spiral seen in \emph{STIS-HST} UV images.  As
explained in the previous section, the spectral analysis has been performed
in two separate regions: the annulus, defined as a ring with inner radius
of $r=2$\farcs5 and outer radius of $r=7$\farcs9, and the unresolved
\emph{Chandra} core, with r$<$1\farcs5. While the first region contains the
individual star forming knots that trace the UV nuclear spiral, the
position of the unresolved \emph{Chandra} core coincides with that of the
UV nuclear star cluster to better than the \emph{Chandra} astrometric
accuracy of 0\farcs6.

\subsection{The unresolved core}

Very recent \emph{STIS-HST} ultraviolet imaging and spectroscopy of the core
region of NGC~4303 have unambiguously identified a compact, massive
and luminous stellar cluster, i.e.  a nuclear super star cluster or SSC
\citep{colina-rosa}.  No further contribution is required to explain its UV
emission, neither further ionizing flux  is needed to account for  the
observed nuclear H$\alpha$   luminosity.   Therefore,  using  only  UV
observations it could not be  inferred whether the NGC~4303 nucleus is
hosting an AGN.   The AGN  classification  is  based on  optical  line
ratios which  have large  uncertainties due  to the weakness  of  some
lines.  The   last revision  of the  NGC~4303   classification is from
\citet{colina-rosa}, using one-dimensional   high-resolution  spectra,
the  result  being  a  [OI]-weak  LINER.   The  ultimate origin of the
activity in LINERs is  still a matter  of discussion. High  energy and
multifrequency data are needed   to check whether the stellar  cluster
itself can account   for the observed   properties,  or an  additional
source of energy  -an accreting black  hole-  is required.  If a black
hole was needed, this could be the  first time, to our knowledge, in
which it would be probed that a  black  hole coexists with a stellar cluster
within the central 3  central pc of a galaxy.   The first step is then
to  establish   the contribution of  the  SSC  to the overall spectral
energy distribution, SED, of the core in NGC~4303.

\subsubsection{X-ray emission associated to the core stellar
  population}
\label{coressc}

The unresolved-core spectrum of NGC~4303 has a clear contribution from
a      thermal       plasma     whose     temperature,        T,    of
$kT=0.65^{+0.24}_{-0.17}$~keV, agrees well with the temperatures found
in  other star-forming regions. In  particular, this temperature
agrees with the temperature found for one of the thermal components of
the annular spectrum, kT$_{\rm I}=0.8^{+0.1}_{-0.2}\,{\rm keV}$. Its
natural explanation is then that it corresponds  to the X-ray emission
of the nuclear star cluster detected in the UV.

\begin{figure}
\includegraphics[angle=-90,scale=0.7]{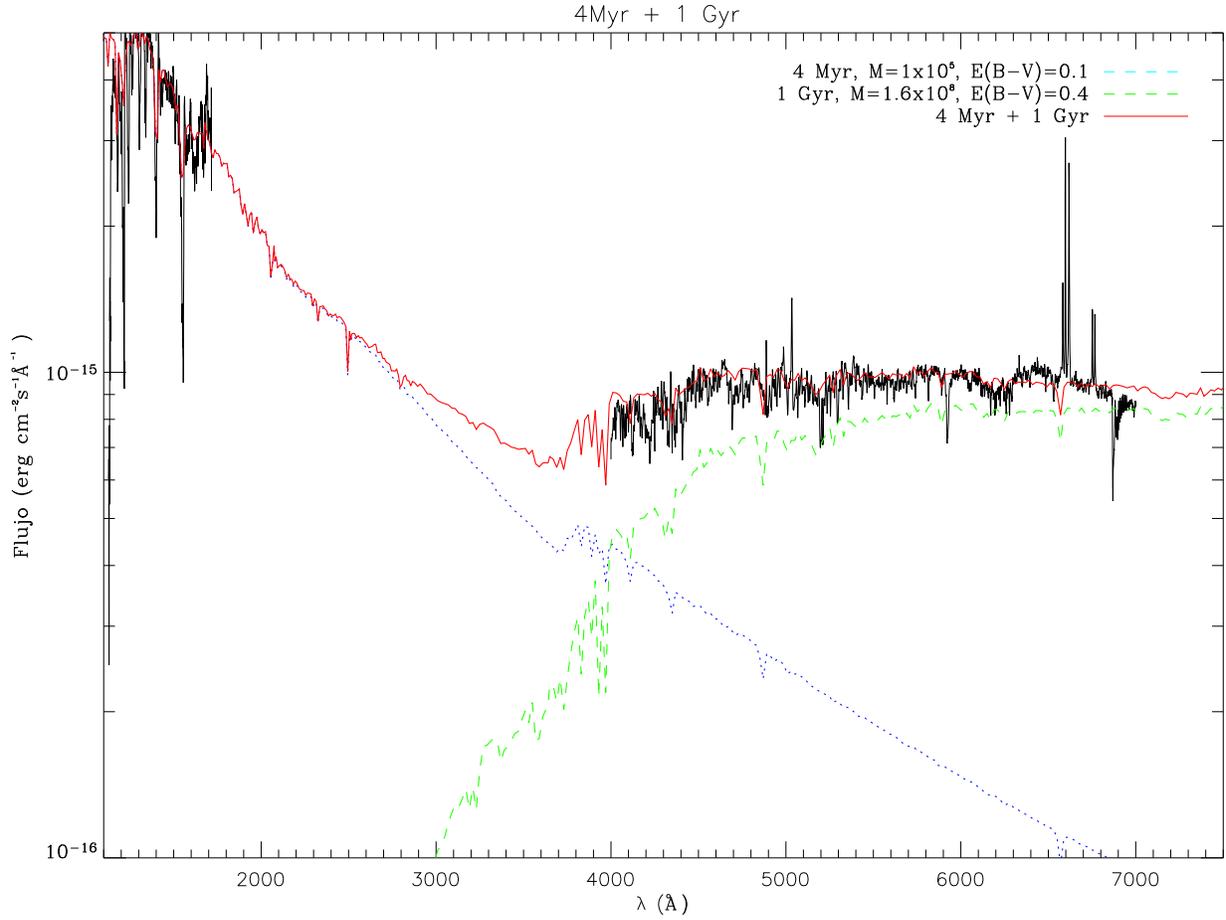}
\caption{Observed ultraviolet to optical spectral energy distributions of the NGC~4303
nuclear region.  The plot shows  the 4 Myr contribution  (dotted line)
expected from the super stellar cluster identified in the core and the
1  Gyr population  (dashed line) from  the disk  and their combination
(solid line). See the text for details, Section~\ref{coressc}.}\label{populations}
\end{figure}

We present in Figure~\ref{populations} the UV to optical spectral energy
distribution of the core of NGC~4303 nuclear region (see
\citet{colina-rosa}).  The UV STIS slit ($0\farcs2$ in diameter) is smaller
than the optical aperture (1\arcsec$\times$1\arcsec), but since the central
UV SSC is not resolved by STIS, we can assume that the STIS slit contains
most of the core UV continuum. The analysis of the UV--optical SED clearly
shows that different populations of stars are co-existing in this region:
the very young super stellar cluster found by \citet{colina-rosa},
dominating the UV continuum, and an older population dominating the optical
range.

According to \citet{colina-rosa}, the core SSC UV spectrum is well
reproduced by a 4~Myr old synthetic instantaneous starburst of solar metallicity
and  with an initial mass of $10^{5} M_{\sun}$ (for a
Salpeter initial mass function, IMF, between 1 and 100 $M_{\sun}$) and obscured
with E(B-V)=0.10 (Large Magellanic Cloud extinction law).  This synthetic
ultraviolet spectrum has been plotted in Figure~\ref{populations} over the
observed one.  It can be seen in the figure that while the SSC completely
dominates the UV continuum of the core region, it provides only a small
fraction of the optical continuum and is negligible at IR wavelengths.

The  optical continuum corresponds to  an  evolved population 1--5 Gyr
old.    We have plotted   in Figure~\ref{populations}  a synthetic SED
corresponding to  a  1 Gyr old  population, with  initial mass  around
$10^{8} M_{\sun}$ (for  a  Salpeter IMF
between 1  and  100\,$M_{\sun}$) and  obscured  with E(B-V)=0.4. The
evolutionary synthesis models by Bruzual  and Charlot have been used to
perform  the analysis   (see  \citet{bruzual}    and  the ftp     site
ftp://gemini.tuc.noao.edu/pub/charlot/bc96). The fact that the   young
cluster is less affected by reddening than the old population seems to
be normal in massive star-forming regions:  the ionizing radiation and
the release  of mechanical energy  by the  massive cluster effectively
'clean up' the surroundings of the  massive stars, blowing the dust to
certain distances  \citep{maiz}.  We can see  in  the figure  that the
combination  of the 4   Myr starburst  and  the  1 Gyr  old population
reproduces very  well  the  UV--optical spectral energy  distribution.
Similar fits can be obtained assuming an  older population up to 5 Gyr
and    different star   formation   scenarios   (instantaneous, medium
duration, continuously decreasing,...), with the  initial mass of this
old population being constrained in any case to the range 1--5 $\times
10^{8}    M_{\sun}$.  The  procedure  shows  that, in any case, an older
population  is dominating  the optical  continuum  within the  central
1\arcsec$\times$1\arcsec.  Nevertheless, this combination of young and
old   population  underestimates     the   observed  near     infrared
emission. \citet{colina-wada} measured an V--H around +3.5 in the core
region, while an evolved population affected by a colour excess around
0.4 would have V--H around +2.5, at most.

This excess in H-band luminosity could be due to the presence of some red
supergiant stars, as discussed in \citet{colina-rosa}.  According to the
predictions of evolutionary synthesis models \citep{cmh,leitherer99}, the
UV central cluster would be too young (around 4 Myr) to host a significant
amount of red supergiants. The presence of red supergiants in a massive
young cluster at solar metallicity peaks at around 10 Myr. This could be a
hint that a previous star formation episode had taken place in the core of
NGC~4303 nucleus some 10 Myr ago. If such a starburst would be present, it
could contribute significantly to the hard X-ray emission, since at this
age there is also a peak in the population of active high mass X-ray
binaries \citep{cervino98tesis}.  The initial mass of such a starburst, as
required to reproduce the H band luminosity, would be around 5 $\times
10^{5} M_{\sun}$.  Such a starburst would produce a strong UV and optical
continuum which has not been detected at all. If a 10 Myr old starburst
were present, it should be completely obscured in the UV and optical
ranges, which does not seem to be consistent with the extinctions derived
from the young and old stellar populations.  We conclude therefore that,
although some red supergiants associated to the young cluster could already
be present, there are no reliable hints of the presence of an intermediate
age starburst around 10 Myr old.

Once the different populations within the core of NGC~4303 nucleus have
been identified, we are able to estimate their associated X-ray
emission. We have used for this purpose the CMHK evolutionary synthesis
models (\citet{cervino98tesis, cervino}\footnote{The models have been
computed using tracks with enhanced mass-loss rates for a self-consistent
comparison with \citet{colina-rosa}} and references therein; 
the results are available in the WWW
server \url{http://www.laeff.esa.es/users/mcs/SED/}), 
which compute the expected X-ray emission (up to 10 keV) produced
during a massive star formation episode, and originated by the diffuse gas
heated by the release of mechanical energy, supernova remnants and binary
systems.  A correction factor of 0.735 has been applied to the model
results for the transformation between the 2--120 M$_\odot$ mass range to
the 1--100 M$_\odot$ mass range used in \citet{colina-rosa}, for a
Salpeter IMF.

We have computed the X-ray luminosity associated to the 4 Myr cluster in
the core of NGC~4303 nucleus. For this we have assumed the properties of
the starburst derived by \citet{colina-rosa} from the UV continuum, as
discussed above. The total UV continuum within the core provides the
absolute normalization for the CMHK models. The total X-ray emission is
parameterized by 2 additional factors: the fraction of stars evolving in
binary systems and the efficiency in conversion of mechanical energy into
X-ray luminosity (by heating the diffuse gas). As discussed in
\citet{cervino}, the fraction of binary stars affects mainly the hard X-ray
emission, while the efficiency parameters constrains basically the soft
X-rays.  

The predictions are compared in Table~\ref{synth_X_ray} with the soft and
hard X-ray luminosities derived in previous sections. It can be seen that
this young cluster could easily account for all the soft X-ray luminosity
detected with \emph{Chandra}, assuming a mechanical energy to X-rays
conversion efficiency of around 15\%.  On the other hand, the observed hard
X-ray emission would be underestimated by the models by around 2 orders of
magnitude. Nevertheless, as discussed above, we can identify 2 different
spectral components in the X-ray emission of NGC~4303 core, a thermal one
well reproduced by a Raymond-Smith model, and a power law component.  We
show in Table~\ref{synth_X_ray} that the evolutionary synthesis model can
reproduce in a consistent way \emph{both} the soft and hard X-ray
luminosity associated to the Raymond-Smith component in the core, assuming
a rather low efficiency of around 3\%. \citet{cervino} found values of this
efficiency between around 10\% and 90\% (after correction from UV continuum
reddening) in a sample of star forming galaxies. CMHK models do not include
the details of the interaction between the high velocity gas released
during a starburst episode and the interstellar medium. In principle,
values of the efficiency between 0\% (no interaction, and therefore no
diffuse gas heating) and 100\% (all the released mechanical heating the gas
and being so converted in X-ray luminosity) would be physically
possible. In the context of the analysis of the starburst--AGN connection,
the CMHK models allow to estimate the maximum X-ray energy that could be
attributed to a starburst episode (i.e., the luminosity obtained assuming a
100\% efficiency). If the observed luminosity is below this value, no
additional energy sources would be needed to explain it. Otherwise, if the
predictions underestimate clearly the observations, even allowing for a
100\% efficiency, this would support strongly the presence of an additional
energy source, not directly related to the starburst process.  We conclude
therefore that the thermal X-ray emission in the core of NGC~4303 is most
likely originated by the starburst episode taking place there.

On the other hand, a starburst episode at 4~Myr would be just starting to
form high-mass X-ray binaries (HMXB).  For a massive binary system to
become active in X-ray emission, the more massive star should have finished
its lifetime, leaving a compact component on which mass from the lower mass
secondary would be able to accrete.  HMXBs would therefore not be present
in the cluster until the more massive stars have exploded as supernovae,
i.e., not before the first 4--5 Myr.  The evolution of HMXBs is included in
the CMHK models \citep{cervino98tesis, cervino}, but our results show that
the X-ray luminosity associated to them will not be enough to account
for the observed hard X-ray emission, whatever the binary fraction is
assumed. Nevertheless, since the formation of the individual most massive
stars, and thus HMXBs, is a highly stochastic process, Poissonian
fluctuations might be very important. In Table~\ref{synth_X_ray} we have
given the 90\% confidence interval for the predicted X-ray luminosities
(\citet{cervinop} describe how the sampling and Poissonian errors have been
considered in the models). It can be seen that the range of luminosities
within this confidence level is very large, so that the upper limit of the
predictions could be marginally consistent with the observations.

We have checked the predictions of the CMHK models with the empirical
calibration of the number of HMXB with $L_x \ge 2\times 10^{38}$ erg
s$^{-1}$ vs. the star formation rate presented by \citet{grimm2002}. From
the H$\alpha$ luminosity within a $1 \arcsec \times 1 \arcsec$ aperture of
1.2$\times 10^{39}$ erg\,s$^{-1}$ \citep{colina-arribas}, we derive a star
formation rate $SFR = 0.013 M_\odot$ yr$^{-1}$, using the semiempirical
calibration of \citet{rosa2002}. The calibration by \citet{grimm2002} gives
then a number of bright HMXB of around 0.04, which we consider
negligible. Therefore, we conclude that the hard component detected in the
core of NGC~4303 nucleus is probably not related to the 4~Myr starburst
episode, although the high degree of uncertainty in these estimates
does not allow us to reject this possibility.

Similarly, we have estimated the total  X-ray luminosity that could be
produced by the expected population of low-mass X-ray binaries (LMXBs)
associated to the old 1--5~Gyr  stellar population. We have done  this
by comparing with the known population of active  LMXBs in our galaxy,
which is  estimated  around  100,  with X-ray  luminosities $L_x  \sim
10^{38}$ erg\,s$^{-1}$ \citep{persic}.  Scaling  the mass of  the  old
population ($\sim  10^{8-9} M_{\sun}$, the exact  value depends on the
extrapolation of the IMF to low masses) with the mass  of stars in the
Galaxy ($\leq 10^{11} M_{\sun}$),  we would  expect  at most  1 active
LMXB, if  any,  associated to   the  old population   in the  core  of
NGC~4303.   We conclude  therefore  that  LMXBs are not   the dominant
contributors to the hard X-ray luminosity in the core of NGC~4303.

More  detailed X-ray spectroscopy    would be needed to identify   the
origin   of this  hard component.    Our  results favor an origin  not
connected with the core SSC, nor with  the old population, although we
cannot completely reject the possibility  that it could be  originated
by some high-luminosity X-ray binaries.

\subsubsection{Does the core of NGC~4303 also harbour an active nucleus?}

The strongest evidence for the  presence of a (hidden) active  nucleus
in  NGC~4303 comes indeed from  the X-ray observations presented here,
as already  advanced in \citet{colina-rosa}.   The hard X-ray emission
arising only from  the unresolved core could well  be due to a  hidden
AGN. However, both its luminosity ($\sim  10 ^{39}$ erg\,s$^{-1}$) and
the  flatness of its spectral shape  (see the discussion below) can be
used   either  as  an  argument or   as   a counter-argument for  that
hypothesis.  In the following we  discuss evidence for and against the
co-existence of a nuclear black hole and the super star cluster in the
nucleus  of  NGC~4303.    We     analyse the   implications     of the
\emph{Chandra} X-ray data   by themselves, and  also   through a wider
analysis of the multiwavelength spectrum of the core of this galaxy.

The hard X-ray tail of the core spectrum is formally fitted by a power
law  with photon index  $\Gamma=1.6\pm0.3$,  which also contributes to
the soft X-ray emission. This spectrum is rather flat when compared to
typical  radio quiet AGN   and  LINERs, i.e.   $\Gamma$ in   the range
1.7--2.3   e.g. (\citet{reeves,george,georganto}     and    references
therein). However, it   is
still formally compatible with 1.9 at 90\% confidence level, given the
large uncertainty.   A fit  with $\Gamma$ fixed  to  1.9, to  test the
possibility of a   highly absorbed AGN emission, gives   statistically
worse results, hence prevents us to take further conclusions.  The low
photons statistics does  not  allow either testing more  sophisticated
models like reflection by either cold or  ionized warm gas.  No counts
are detected at the energy of the iron K$\alpha$ emission.  No sign of
variability has been detected in the  energy range between 1.5--10 keV
during the 28600   s \emph{Chandra} observation   elapsed integration
time.

The observed hard X-ray luminosity, after correction for line-of-sight
absorption, $L_{2-10\,keV}=8^{+3}_{-2} \times 10^{38}$\,erg\,s$^{-1}$,
is quite low for Seyfert 1 galaxies or  quasars, that typically exceed
$10^{43}$\,erg\,s$^{-1}$. Nevertheless, similar luminosities have been
measured for a  few  low  luminosity AGN  or  LINERS.  The soft  X-ray
luminosities  of LINERs as  measured  with  ROSAT  are  in the   range
$10^{38-41}$ erg\,s$^{-1}$  \citep{komossa}.  The   \emph{ASCA}, 2--10
keV, luminosity   of  a sample  of  21  LINERs  and  17 low-luminosity
Seyferts       range   from   $4\times10^{39}$  to    $5\times10^{41}$
erg\,s$^{-1}$,  with part of   the  flux coming  from  a  soft thermal
component  and  part from  a   power  law with  $\Gamma\approx  1.8  $
\citep{terashima}.  The lower end of the X-ray luminosity distribution
is $2\times10^{39}$ erg\,s$^{-1}$  if only the  power law contribution
is  considered.  \emph{BeppoSAX} observations  of  6 type-2 LINER  and
transition galaxies show 2--10 keV luminosities in the range $10^{39}$
to $10^{41}$   erg\,s$^{-1}$  \citep{georganto}.  \emph{Chandra}   has
detected  weak X-ray sources in  the nuclei of  some Seyfert galaxies,
LINERs  and LINER/H~II transition objects  \citep{ho}.   The 2--10 keV
luminosities  of   their  detected sources  range  from   $10^{38}$ to
$10^{41}$   erg\,s$^{-1}$  and most   of    the non-detected ones  are
transition type objects.    The  median 2-10\,keV luminosity   of  the
objects which have clear classification  as either LINER or Seyfert in
the \citet{ho} sample  is   $1.2 \times 10^{40}$\,erg\,s$^{-1}$.   The
median  luminosity   of the   clear \citet{ho} LINERs    is  $7 \times
10^{39}$\,erg\,s$^{-1}$ and  of  the  Seyferts in   their sample  $1.6
\times 10^{40}$\,erg\,s$^{-1}$. The core  luminosity of NGC 4303 is at
the lower   end of  the   above  distributions,  compatible  with  low
luminosity Seyfert or LINERs.  The     black hole mass requested    to
explain the observed X-ray luminosity is quite low for an AGN, only of
$\sim  1.2\times  10^4$ M$_\odot$.  This mass  is   inferred using the
bolometric correction ($L_{bol}/L_X$) of  10  and an Eddington   ratio
($L_{bol}/L_{Edd}$) of 0.01,   both quantities considered  typical for
low luminosity AGN \citep{awaki}, together with the luminosity emitted
by a black hole accreting at its Eddington luminosity: $L_{Edd}
\simeq 1.3 \times 10^{38} M_{BH}/M_\odot$.  An alternative possibility is
that a higher mass  black hole, with  mass $\sim  10^{6-7}$ M$_\odot$,
similar to known  Seyfert nuclei, does exist in  the core of NGC~4303,
but accreting with low  radiative efficiency. Indeed, the  flatness of
the X-ray spectrum, $\Gamma  \sim 1.6$, suggests the  possibility that
the nuclear emission could be due to the so-called Advection Dominated
Accretion Flows, ADAF,
\citep{rees,narayan,fabian}. In this scenario, the hard
X-ray component would be dominated by a bremsstrahlung thermal emission
from a population of $\sim 100$\,keV electrons which is therefore much
harder than typically observed in Seyfert galaxies. This kind of accretion
has been suggested as the origin of the hard X-ray emission detected in the
elliptical galaxy NGC~1052, characterized by a flat spectrum with $\Gamma
\sim 1.4$ \citep{guainazzi}. As shown above, a fit to the hard X-rays in
NGC~4303 that includes   a  bremsstrahlung model gives   an acceptable
result.  However,  the best   fit  bremsstrahlung  temperature,  $\sim
8^{+60}_{-5}$\,keV,  is  lower than   the  predictions  of  the   ADAF
scenario,  although  the  uncertainties in  the  temperature  make  it
compatible   with    a   higher   value,    70\,keV,   within    90\%
confidence. Therefore an ADAF is not favoured by  the current data but
cannot be discarded.  The presence   of spectral components with  high
absorption (N$_{H} \ge 10^{24}$ cm$^{-2}$)  cannot be excluded either,
but there is at  present no clear  evidence for it  as no good fit was
achieved with the current \emph {Chandra} data.

The data presented  until now are  not enough to elucidate between AGN
or starburst origin  of  the hard   X-ray core emission  in NGC  4303.
Multifrequency information is therefore needed.

A   formal  comparison of  the   UV-optical   luminosity to the  X-ray
luminosity can be made through the so-called optical-to-X-ray spectral
index
$\alpha_{ox}=-\log\frac{F_{\nu}(2\,keV)}{F_{\nu}(2500\textrm{\AA})}/\log\frac{\nu(2\,keV)}{\nu(2500\textrm{\AA})}$.
Using  the UV emission at  2500~\AA~from \citet{colina-rosa} and the 2
keV core   emission from    this paper, we     obtain $\alpha_{ox}\sim
2.0$.  The  mean value  of  $\alpha_{ox}$ in  radio quiet  AGN is 1.48
\citep{laor},   with  typical  values   in   the   range 1.4  to  1.65
\citep{green}.  The value  obtained  for  the  core  in NGC  4303   is
therefore  quite large, indicative of  relatively  weak X-ray emission
compared to radio  quiet  AGN, either due to   high absorption in  the
X-rays only or to a different origin of  one or two of the components.
This result is, of course, not surprising as
\citet{colina-rosa} have already shown that the UV flux in the core of NGC
4303 comes from starburst  emission only. Another  possible comparison
is the 2--10 keV X-ray to H$\alpha$  luminosity ratio. There is a good
correlation between both quantities in Seyfert  1 galaxies and quasars
that has  been shown to extend down  to low luminosity AGNs (\citet[][
and references  therein]{ho} ).  This  is consistent  with the optical
lines being explained     by photoionization  from   the  central  AGN
continuum in both high and  low luminosity AGNs.  \citet{ho}, however,
find  that   in some   narrow-line  objects in  their   sample  of low
luminosity AGN, the $L_X/L_{H\alpha}$ ratio is  a factor 10 lower than
expected     by  the   above correlation.     We     have measured the
$L_X/L_{H\alpha}$ ratio  in the core of NGC   4303 using the H$\alpha$
luminosity   within a  $1   \arcsec \times   1   \arcsec$ aperture  of
1.2$\times 10^{39}$erg\,s$^{-1}$ \citep{colina-arribas} and the 2--10
keV flux  in this paper. We obtain   a value of  $L_X/L_{H\alpha} \sim
0.8$, a  factor 8 lower than  expected from the  best  fit for  type 1
objects in  \citet{ho}, and also lower than  their mean value  for the
type 2 objects in  their sample with median $L_X/L_{H\alpha}\approx 2$
and a large scatter. This result can be interpreted as  the need of an
additional ionizing source other   than the X-ray emitting  source  to
produce the observed H$\alpha$  luminosity. This result is, again, not
surprising, given that \citet{colina-rosa}  explain that the  observed
UV flux from the nuclear starburst provides enough ionizing photons to
account for the observed H$\alpha$ emission.

\subsubsection{Can the hard X-ray emission be attributed to other sources?}

As shown in Sec.~\ref{coressc}, according  to our current knowledge of
the stellar evolution, there is only a  marginal probability that in a
very young cluster, as   the one in  the  core of NGC~4303, high  mass
X-ray binaries have formed.

Nevertheless, non-nuclear point sources with similar hard X-ray
luminosities have been detected in some X-ray images of nearby
galaxies. Examples are the two brightest point sources in NGC~253 with mean
luminosities of 9.5 and $5\times 10^{38}$erg\,s$^{-1}$, and flux variations
of factor $\sim 2$ \citep{pietsch} and the most luminous X-ray source in
M82 which is highly variable with a peak X-ray luminosity $\sim 9\times
10^{40}$\,erg\,s$^{-1}$ \citep{kaaret}. Recently, \citet{foschini} have
performed a study of ultraluminous X-rays sources in a sample of 10 nearby
Seyfert galaxies. They have found 18 such sources, 10 of which have
2-10\,keV luminosities similar to or larger than $10^{38}$erg\,s$^{-1}$, in
some cases providing a larger contribution to the overall X-ray flux of the
host galaxy than the active nucleus itself.  The most plausible explanation
for these bright sources according to \citet{foschini} is that they are
accreting black holes X-ray binaries. If one of such sources is in the core
of NGC~4303, in order to account for the hard X-ray emission detected with
\emph {Chandra}, the black hole component in the binary system should have
a mass of $\sim 10$ M$_\odot$, assuming it is radiating at its Eddington
limit and isotropically. The black hole mass should be higher if the ratio
of the bolometric to the hard X-ray luminosity is greater than 1, e.g.  it
is $\ge 10$ in AGN, or if the accretion is sub-Eddington.

The origin of ULX is nevertheless controversial, with different options
listed, e.g.  by \citet{taniguchi} and commented here.  Most of the ULX are
located in regions of active star formation suggesting that there might be
a physical link between both phenomena.  It has already been suggested that
those that are located near the dynamical centre of the host galaxy can be
low luminosity AGN.  This possibility for the NGC~4303 core has already
been discussed in the previous subsection.  Alternatives are related with
very luminous supernova remnants expanding in a very dense medium and with
accretion-powered binaries. In the latter case the compact component can
either be similar to Galactic black hole candidates with masses up
10\,M$_\odot$ or a new population of $10^2-10^4$ intermediate-mass black
holes \citep{colbert}. Most of the ULX detected so far are best explained
in the context of accreting black holes, mainly because of the detection of
flux variations in some of them and the fact that observed spectra are
usually well fitted with multicolor disk blackbody emission and/or a power
law as is the case for Galactic black holes.  The unusually high innermost
disk temperatures, 1.2--1.8\,keV, are explained with black hole rotation
\citep{makishima}.  There remains the question on whether the mass is
similar to that in ordinary stellar mass black holes or whether it is
higher, up to $\sim 10 ^4\,{\rm M}_\odot$. The first case requires that the
binary system is observed in an epoch of unusually high, even
supercritical, accretion rate, and/or with a preferred orientation assuming
beamed radiation \citep{king, zezas, roberts}. In the second case, the main
problem is how such intermediate-mass black holes can be formed.
\citet{taniguchi} propose that they are formed in very dense circumnuclear
regions of galaxies where several hundreds of massive stars have
evolved. The continuous merging of these compact remnants in $\sim 10^9$
years can led to the formation of intermediate-mass black holes.
Alternative models proposed by \citet{taniguchi} are interaction with a
satellite galaxy or gas accretion onto a seed black hole.

We discuss here the possibility that the hard X-ray emission, in the core
of NGC~4303 is an ULX due to one of such sources.  Although we have
discussed in Sect.~\ref{score} that models including multicolor disks do not
reproduce satisfactorily the observations, \citet{strickland} and
\citet{roberts} show that the analysis of ULX \emph {Chandra} spectra
favors  simple power laws  with index  spaned in  a range  of 1.8-2.9,
which would be consistent with the index of the  power law of the best
fit model for the  core region, $\Gamma=1.6\pm0.3$.  The position  of
the ULX is coincident  with the stellar   cluster located in  the very
centre of the galaxy, but as has been discussed in Sec.~\ref{coressc},
the cluster is  too  young and  even  the most massive stars   are not
expected to  have  exploded  as  supernovae.   Therefore  it  is  very
unlikely that the origin of the ULX is from  the young stellar cluster
itself.  There  remains still the possibility that  the  black hole is
the remnant from a supernova explosion from the 1-5 Gyr old population
also  present in  the nucleus of  the  galaxy (Sec.~\ref{coressc}). If
this $10^8\,{\rm M}_\odot$  population contained a  few  stars of more
than $\sim  30\,{\rm   M}_\odot$,  they should have   completed  their
evolution in the first $\sim 10^7$ yr and  left a few compact remnants
of at least  2\,M$_\odot$. Then it is   possible that at  least one of
them has experienced disk accretion in  the remaining life time of the
cluster and increased its  mass to $\ge  10{\rm M}_\odot$. The  strong
star  formation  activity in the  nuclear  region of NGC~4303 suggests
that  there is enough  gas to feed the black   hole accretion. One can
also  speculate  that  intermediate-mass black  holes  could have been
formed in one of the stellar clusters  in the vicinity of the NGC~4303
core, following the  mechanism suggested by \citet{taniguchi} provided
there is at least one cluster  that is old and  dense enough, and sink
down to the core via dynamical friction.

We can therefore conclude that, even if there is no definite evidence for
it, the hard X-ray source in the core of NGC~4303 could be similar to the
ULX found in other galaxies and explained in most of the cases with
accreting binaries, in which the compact object is either a $\sim 10\,{\rm
M}_\odot$ or an intermediate-mass black hole.

\subsection{The annular region}

We have also applied the CMHK models to the annular region.
\citet{colina-rosa} found that the brightest circumnuclear star-forming
knot could be fitted by a $8\times 10^3$ M$_\odot$ 3.5~Myr old starburst,
affected by an internal absorption with E(B-V) around 0.1 (Large Magellanic
Cloud extinction law).  As a first approximation, we have assumed a similar
evolutionary stage for all circumnuclear starburst, and have normalized the
UV luminosity to the overall luminosity at 2117 \AA\ within an annulus with
internal and external radii of 0\farcs5 and 4\arcsec, respectively, as measured by
\citet{colina-vargas}.  This region contains indeed most of the UV
continuum luminosity, as can be appreciated in Figure~\ref{hst_chandra}.

We compare in Table~\ref{synth_X_ray} the observed soft and hard X-ray
luminosities of this annular region, as measured on \emph{Chandra} images,
with the predictions from the CMHK models for a young 3.5~Myr star
formation episode having transformed 5 $\times 10^5\,M_{\sun}$ into
stars. It can be seen that the emission associated to these star formation
knots accounts for both the soft and hard X-ray emission observed with
\emph{Chandra}, assuming a mechanical energy to X-rays conversion
efficiency of around 30\% . While the model predictions seem to underestimate the hard X-ray
emission by a factor around 2.5, we consider the discrepancy negligible when
compared to the accuracy of both the hard X-ray measurements and the model
estimates (the observed hard X-ray luminosity given in
Table~\ref{synth_X_ray} comes from the model extrapolation of the thermal
components detected in soft X-rays).  No additional components would be
required to explain the X-ray emission from the annulus. Moreover, the fact
that there are no high-luminosity binaries associated to the circumnuclear
starbursts, although the total mass and age is similar to that of the core
SSC, supports our previous conclusion that the hard X-ray core component is not associated
to the core starburst, either.

We want finally to stress that most of the soft X-ray emission in the
nuclear region of NGC~4303 (including both the annulus and the core) seems
therefore to be associated with star formation episodes. As concluded by
\citet{cervino}, this could be the case for many low luminosity active
galaxies, especially for a significant fraction of Seyfert~2 galaxies. The
presence of a hidden active nucleus in these galaxies could therefore be
inferred only from its emission at higher energies, or from its indirect
effects on the optical emission line spectra.  \citet{n1068} show indeed
that even in the case of the prototype of Seyfert~2 galaxy, NGC~1068, both
the UV continuum and the soft X-ray emission seem to be dominated (or at
least contributed with a significant fraction) by compact star forming
episodes within less than 10 pc from its core. \citet{georganto} also
conclude that the bulk of the nuclear X-ray emission in the low luminosity
AGN NGC~3627 and NGC~5195 originates in their (circum-)nuclear star-forming
regions.

\section{Conclusions}

\emph{Chandra} X-ray images of the low luminosity AGN NGC~4303 have allowed to
resolve the X-ray emission originated in the core ($r\leq1$\farcs5) from
the surrounding annular region ($2\farcs5\leq r\leq 7$\farcs9). The
astrometry of the \emph{Chandra} core coincides to less than
0\farcs3$\pm0.6$ with the position of the ultraviolet  core. While the
X-ray emission of the annular region is well reproduced assuming 2 thermal
Raymond-Smith components, the best fit for the core region requires both
a thermal and a power law component. Most of the overall soft X-ray
luminosity from the nucleus of NGC~4303 is emitted by the surrounding
annular region (70\%), with a minor fraction originating within the
core. On the other hand, more than 90\% of the hard X-ray nuclear 
emission is produced within the unresolved core. 

Our results show  that most of the soft  X-ray energy from the central
region   of   NGC~4303   is   associated   to    massive  star-forming
episodes. Evolutionary  synthesis models show  that the total soft and
hard X-ray emission originated in the annular  region is associated to
the massive  star-forming knots  identified there  by \emph{STIS-HST}  UV
imaging and spectroscopy. Similarly, the young massive stellar cluster
identified within  the core region  can also  account for the emission
from the Raymond-Smith   core component.  On  the  other hand,
evolutionary  synthesis models underestimate  by at  least one order of
magnitude the core  hard X-ray emission which  is modeled  by a rather
flat   power  law.   It  should  be noted,   however,  that the  large
uncertainties, both in the measurements  and in the predictions of the
evolutionary synthesis model hard X-ray values, make both luminosities
marginally compatible at the 90\% confidence level.

The most striking result is the detection of the hard X-ray tail in the
core emission of NGC 4303, with a luminosity of $\sim 10^{39}$
erg\,s$^{-1}$, intermediate between the expected values in standard AGN and
pure starburst galaxies and normal for LINERs. The results of the
evolutionary synthesis models do not allow to exclude the presence of a few
high-luminosity black-hole binaries within the core (the predictions
within a 90\% confidence level would be marginally consistent with the
observations), but the models show that this is not a very likely
scenario.  We have to look for other alternatives. One obvious solution for
the origin of the hard X-rays is the presence of a hidden low luminosity
AGN. It has been discussed that the observed properties (luminosity,
spectral shape, X-ray to H$\alpha$ luminosity ratios) are similar to some
type-2 low luminosity LINERs or even low luminosity Seyfert 2, like those
in the \citet{ho} sample.  We have shown that the observations are just
compatible with different AGN scenarios: a highly absorbed (N$_{H} \ge
10^{24}$ cm$^{-2}$) nuclear emission, a low mass AGN (M$_{BH} \sim 10^{4-5}
{\rm M}_\odot$) or a higher mass black hole, but accreting with very low
radiative efficiency like in the ADAF scenario.  If NGC 4303 indeed
harbours a low luminosity AGN, it would be the first one to our knowledge
where an additional source, a super stellar cluster, has been detected in
the very nucleus and co-exists with the AGN in the central 3 pc.

The  alternative to a  low luminosity AGN is  that the nuclear stellar
cluster is  the only ionizing source, while  the hard X-ray luminosity
comes from  a source  similar  to the  ultraluminous X-ray  point-like
off-nuclear sources, ULX, detected in a number of nearby galaxies. The
suggested  origin for such ULX sources  is accretion onto a black hole
binary,  which  for NGC~4303  requires a black  hole mass  of at least
$\sim 10$ M$_\odot$.  In this context  it is interesting to note that
recently, \citet{weaver} speculated with the idea  that such ULX black
holes   may    be  the  precursors  to   AGN   activity.  According to
\citet{weaver} (and   the references she  mentions),  if born in dense
star clusters near the centers of galaxies, ULX black holes could sink
to the  core via   dynamical  friction,  eventually growing   into   a
supermassive black hole.   In  NGC 4303  the dense  stellar cluster is
already at the core of the  galaxy and an ULX might  also be there and
ready to  grow  fed by the starburst   itself. Are  we looking  at the
nucleus  of a  galaxy  in  the  phase  preceding  the birth  of  a
'normal'-luminosity AGN powered by   a supermassive black hole  ? Only
better signal to noise  observations  of its high-energy  emission and
its  indirect  effects  on  the  emission lines  could provide further
clues.

\begin{acknowledgements}

EJB, JMMH and MC have been partially supported by Spanish grant
AYA2001-3939-C03-02. 
LC has been partially supported by Spanish grants
PB98-0340-C02 and AYA2002-01055. We acknowledge interesting suggestions 
from an anonymous referee. 

\end{acknowledgements}

\clearpage

\clearpage

\clearpage

\clearpage

\clearpage

\clearpage

\clearpage

\clearpage

\thispagestyle{empty}
\begin{deluxetable}{c l l l l l l l l l l l l}
\rotate
\tabletypesize{\scriptsize}
\tablecaption{Locations, count rates and luminosities of the point-like 
sources detected in the NGC~4303 \emph{Chandra} image \label{table_sources}}
\tablewidth{0pt}
\tablehead{
\colhead{{\bf Source}} & \colhead{{\bf RA}}& \colhead{{\bf DEC }}& \colhead{{\bf  C$_{soft}$}} & \colhead{{\bf  CR$_{soft}$}} & \colhead{{\bf L$^1_{soft}$}}& \colhead{{\bf L$^2_{soft}$}}& \colhead{{\bf L$^3_{soft}$}}& \colhead{{\bf  C$_{hard}$}} & \colhead{{\bf  CR$_{hard}$}} & \colhead{{\bf L$^1_{hard}$}}& \colhead{{\bf
L$^2_{hard}$}}& \colhead{{\bf L$^3_{hard}$}}\\
\colhead{ } & \colhead{(J2000)} & \colhead{(J2000)} & \colhead{} &\colhead{$10^{-4}\,s^{-1}$} & \colhead{$10^{38}\,erg s^{-1}$}&  \colhead{$10^{38}\,erg s^{-1}$}&  \colhead{$10^{38}\,erg
s^{-1}$}& \colhead{}& \colhead{$10^{-4}\,s^{-1}$} & \colhead{$10^{38}\,erg s^{-1}$}& \colhead{$10^{38}\,erg s^{-1}$}
& \colhead{$10^{37}\,erg s^{-1}$}
}
\startdata

a$^*$ &12 $^h 21 ^m 44\fs4 $ & 4 $\degr 29 \arcmin 10\farcs7 $  &  24$\pm$ 6 &  8.6$\pm$ 2.2 &  1.3$^{+ 0.6}_{- 1.4}$ &  0.8$\pm$0.2 & 1.2$\pm$0.3  &   9$\pm$ 4 &  3.3$\pm$ 1.6 & 0.8$^{+ 0.2}_{- 0.2}$  & 0.5$\pm$0.2  & 0.6$\pm$ 0.3 \\
b$^*$ &12 $^h 21 ^m 46\fs7 $ & 4 $\degr 29 \arcmin 59\farcs9 $  &  85$\pm$10 & 30.4$\pm$ 3.7 &  4.7$^{+ 1.5}_{- 4.3}$ &  2.8$\pm$0.3 & 4.1$\pm$0.5  &  21$\pm$ 5 &  7.6$\pm$ 2.1 & 1.9$^{+ 0.6}_{- 0.5}$  & 1.1$\pm$0.3  & 1.3$\pm$ 0.4 \\
c$^*$ &12 $^h 21 ^m 51\fs6 $ & 4 $\degr 31 \arcmin  8\farcs2 $  &  35$\pm$ 7 & 12.6$\pm$ 2.6 &  1.9$^{+ 0.8}_{- 2.0}$ &  1.2$\pm$0.2 & 1.7$\pm$0.3  &   7$\pm$ 4 &  2.6$\pm$ 1.5 & 0.6$^{+ 0.2}_{- 0.2}$  & 0.4$\pm$0.2  & 0.4$\pm$ 0.3 \\
d$^*$ &12 $^h 21 ^m 51\fs6 $ & 4 $\degr 30 \arcmin  6\farcs3 $  &
12$\pm$ 4 &  4.4$\pm$ 1.7 &  0.7$^{+ 0.4}_{- 0.8}$ &  0.4$\pm$0.2 &
0.6$\pm$0.2  &  -&-&- & - & -\\
e$^+${\bf (E,12)} &12 $^h 21 ^m 51\fs9 $ & 4 $\degr 28 \arcmin  2\farcs9 $  &  85$\pm$10 & 30.4$\pm$ 3.8 &  4.7$^{+ 1.5}_{- 4.3}$ &  2.8$\pm$0.3 & 4.1$\pm$0.5  &  27$\pm$ 6 &  9.8$\pm$ 2.3 & 2.5$^{+ 0.7}_{- 0.6}$  & 1.4$\pm$0.3  & 1.7$\pm$ 0.4 \\
f$^+${\bf (E,12)} &12 $^h 21 ^m 52\fs1 $ & 4 $\degr 28 \arcmin 11\farcs2 $  &  62$\pm$ 9 & 22.1$\pm$ 3.3 &  3.4$^{+ 1.2}_{- 3.3}$ &  2.0$\pm$0.3 & 3.0$\pm$0.4  &  38$\pm$ 7 & 13.7$\pm$ 2.7 & 3.4$^{+ 1.0}_{- 0.8}$  & 2.0$\pm$0.4  & 2.4$\pm$ 0.5 \\
g$^+${\bf (A)} &12 $^h 21 ^m 54\fs5 $ & 4 $\degr 29 \arcmin 33\farcs9 $  &  88$\pm$10 & 31.4$\pm$ 3.8 &  4.9$^{+ 1.6}_{- 4.5}$ &  2.9$\pm$0.3 & 4.2$\pm$0.5  &   9$\pm$ 4 &  3.4$\pm$ 1.6 & 0.8$^{+ 0.3}_{- 0.2}$  & 0.5$\pm$0.2  & 0.6$\pm$ 0.3 \\
h$^+${\bf (C)} &12 $^h 21 ^m 55\fs5 $ & 4 $\degr 28 \arcmin 58\farcs5 $  &  96$\pm$11 & 34.3$\pm$ 4.0 &  5.3$^{+ 1.7}_{- 4.9}$ &  3.2$\pm$0.4 & 4.6$\pm$0.5  &  33$\pm$ 7 & 11.9$\pm$ 2.5 & 3.0$^{+ 0.9}_{- 0.7}$  & 1.7$\pm$0.4  & 2.0$\pm$ 0.4 \\
i$^+${\bf (C)} &12 $^h 21 ^m 55\fs5 $ & 4 $\degr 29 \arcmin 10\farcs6 $  &  31$\pm$ 7 & 11.1$\pm$ 2.5 &  1.7$^{+ 0.7}_{- 1.8}$ &  1.0$\pm$0.2 & 1.5$\pm$0.3  &   4$\pm$ 3 &  1.6$\pm$ 1.4 & 0.4$^{+ 0.1}_{- 0.1}$  & 0.2$\pm$0.2  & 0.3$\pm$ 0.2 \\
j$^+${\bf (B,9)} &12 $^h 21 ^m 56\fs6 $ & 4 $\degr 29 \arcmin 24\farcs6 $  & 112$\pm$11 & 40.0$\pm$ 4.2 &  6.2$^{+ 1.9}_{- 5.6}$ &  3.7$\pm$0.4 & 5.4$\pm$0.6  &  10$\pm$ 4 &  3.7$\pm$ 1.7 & 0.9$^{+ 0.3}_{- 0.2}$  & 0.5$\pm$0.2  & 0.6$\pm$ 0.3 \\
k$^+$ &12 $^h 21 ^m 57\fs5 $ & 4 $\degr 29 \arcmin  2\farcs9 $  &  18$\pm$ 5 &  6.4$\pm$ 2.1 &  1.0$^{+ 0.5}_{- 1.1}$ &  0.6$\pm$0.2 & 0.9$\pm$0.3  &  22$\pm$ 6 &  8.0$\pm$ 2.2 & 2.0$^{+ 0.6}_{- 0.5}$  & 1.2$\pm$0.3  & 1.4$\pm$ 0.4 \\
l$^{\dagger}$  &12 $^h 21 ^m 58\fs2 $ & 4 $\degr 30 \arcmin 53\farcs3 $  &  26$\pm$ 6 &  9.4$\pm$ 2.3 &  1.5$^{+ 0.6}_{- 1.5}$ &  0.9$\pm$0.2 & 1.3$\pm$0.3  &  10$\pm$ 4 &  3.8$\pm$ 1.7 & 0.9$^{+ 0.3}_{- 0.2}$  & 0.5$\pm$0.2  & 0.6$\pm$ 0.3 \\
m$^+${\bf (D,11)} &12 $^h 21 ^m 58\fs3 $ & 4 $\degr 28 \arcmin 12\farcs3 $  & 125$\pm$12 & 44.6$\pm$ 4.4 &  6.9$^{+ 2.1}_{- 6.2}$ &  4.1$\pm$0.4 & 6.0$\pm$0.6  &  19$\pm$ 5 &  6.9$\pm$ 2.1 & 1.7$^{+ 0.5}_{- 0.4}$  & 1.0$\pm$0.3  & 1.2$\pm$ 0.4 \\
o$^{\dagger}$ &12 $^h 22 ^m  0\fs5 $ & 4 $\degr 31 \arcmin 40\farcs4 $  & 124$\pm$12 & 44.4$\pm$ 4.4 &  6.9$^{+ 2.1}_{- 6.2}$ &  4.1$\pm$0.4 & 6.0$\pm$0.6  &  18$\pm$ 5 &  6.6$\pm$ 2.1 & 1.7$^{+ 0.5}_{- 0.4}$  & 0.9$\pm$0.3  & 1.1$\pm$ 0.4 \\
p$^{\dagger\dagger}${\bf (8)} &12 $^h 22 ^m  1\fs3 $ & 4 $\degr 29 \arcmin 37\farcs6 $  & 180$\pm$14 & 64.4$\pm$ 5.2 & 10.0$^{+ 2.8}_{- 8.8}$ &  5.9$\pm$0.5 & 8.7$\pm$0.7  &  34$\pm$ 7 & 12.3$\pm$ 2.6 & 3.1$^{+ 0.9}_{- 0.7}$  & 1.8$\pm$0.4  & 2.1$\pm$ 0.4 \\
q$^{\dagger\dagger}$ &12 $^h 22 ^m  1\fs3 $ & 4 $\degr 29 \arcmin  4\farcs7 $  &  41$\pm$ 7 & 14.8$\pm$ 2.7 &  2.3$^{+ 0.9}_{- 2.3}$ &  1.4$\pm$0.2 & 2.0$\pm$0.4  &  16$\pm$ 5 &  5.9$\pm$ 2.0 & 1.5$^{+ 0.4}_{- 0.4}$  & 0.8$\pm$0.3  & 1.0$\pm$ 0.3 \\
r$^{\dagger\dagger}$ &12 $^h 22 ^m  2\fs5 $ & 4 $\degr 28 \arcmin  5\farcs5 $  &  11$\pm$ 4 &  4.1$\pm$ 1.7 &  0.6$^{+ 0.4}_{- 0.8}$ &  0.4$\pm$0.2 & 0.6$\pm$0.2  &  11$\pm$ 4 &  4.1$\pm$ 1.8 & 1.0$^{+ 0.3}_{- 0.2}$  & 0.6$\pm$0.3  & 0.7$\pm$ 0.3 \\
\tableline
\enddata

\tablecomments{The coincidence, if any, with a known \emph {ROSAT} source is
also shown in brackets following the nomenclature in
\citet{tschoke}.
The  \emph{soft} band corresponds to  0.23-2~keV  and the \emph {hard}
band to 2-10~keV. The luminosities,  L$^1$, have been calculated using
a power law  index of 1.2$\pm$0.2  for  the 2-10~keV band, the  errors
have been derived from the uncertainty in the power law index; for the
0.23-2~keV  band, the  luminosities have  been  derived  from the ECF,
calculated through the  spectral analysis of the  {\bf m} source.  The
luminosities marked with L$^2$  and  L$^3$ have been  calculated using
power    law    index   of  1.6    and    2.9,   respectively,    (see
Section~\ref{off-nuclear} for details). The  errors  have been  derived
from the count  rate errors. The  net  counts of these  sources marked
with $^*$ have been calculated using background 1, $^+$ for background
2,  $^{\dagger}$    background 3 and   $^{\dagger\dagger}$  background
4.  Luminosities have been estimated  in a different  way for the soft
and hard band, see the text for details. }
\end{deluxetable}
\clearpage

\begin{deluxetable}{l l l l l l l l l} 
\tablewidth{0pt}
\tabletypesize{\scriptsize}
\rotate
\tablecaption{ Values for the parameters of the fit models for the core
region.\label{table_core}}
\tablehead{
\colhead{\emph{\bf Model}} & \colhead{\emph{\bf Galactic}} &
\colhead{\emph{\bf Component I}} & \colhead{\emph{\bf Component I}} &
\colhead{\emph{\bf Component I}} & \colhead{\emph{\bf Component II}}&
\colhead{\emph{\bf Component II}}& \colhead{ $\chi^2_{\nu}$} & \colhead{\emph{\bf dof}}  \\
\colhead{} & \colhead{n$_H$} & \colhead{ n$_H$} &\colhead{$\Gamma$/kT/T$_{in}$} & \colhead{Norm } & \colhead{kT} &
\colhead{Norm } & \\
\colhead{(1)} & \colhead{(2)} & \colhead{(3)} & \colhead{(4)} &
\colhead{(5)} & \colhead{(6)} & \colhead{(7)}\\
}\startdata phabs(pwlw) &  1.67$\times$  10$^{20}$(frozen) &   & 1.8 &
7.2$\times10^{-6}$ &  &  & 2.4 & 6\\

phabs(ray) & 1.67$\times$ 10$^{20}$(frozen) &  & 5 & 2.5$\times10^{-5}$ &  &  & 3.3 & 6\\

{\bf phabs(pwlw+ray) }& 1.67$\times$ 10$^{20}$(frozen) &  &$1.6\pm0.3$ &
$5\pm2 \times 10^{-6}$ & $0.65^{+0.24}_{-0.17}$ & $2\pm1 \times
10^{-6}$ & 1.1 & 4\\

phabs(ray+bremss) & 1.67$\times$ 10$^{20}$(frozen) &  &
$0.63^{+0.15}_{-0.26}$ & $1.5^{+0.6}_{-0.8}\times10^{-6}$ & 
$8^{+62}_{-4}$ & $7.1^{+1.5}_{-1.6}\times10^{-6}$ & 0.99 & 4\\

phabs(phabs*pwlw+ray) & 1.67$\times$ 10$^{20}$(frozen)  &
$1.0^{+1.4}_{-0.8}\times$ 10$^{20}$ & 1.9(fixed) & $7^{+3}_{-2}\times
10^{-6}$ & $0.65^{+0.24}_{-0.17}$ & $2\pm1 \times 10^{-6}$ & 1.34 & 4 \\

phabs(mdc) & 1.67$\times$ 10$^{20}$(frozen) &  & 0.9 & $3\times
10^{-3}$ &  &  & 3.9 & 6 \\

phabs(pwlw+mcd) & 1.67$\times$ 10$^{20}$(frozen) &  & 1.9 & $6\times
10^{-6}$ & 1.4 & $1\times 10^{-4}$ & 3.5 & 4\\
\enddata
\tablecomments{
Col. (1)- Spectral models: \emph{phabs:} Photoelectric Absorption,
\emph{pwlw:} Power Law, \emph{ray:} Raymond-Smith, \emph{bremss:}
Thermal Bremsstrahlung, \emph{mcd:} Multi-color disk\\
Col. (2)- Column density in units of cm$^{-2}$\\
Col. (3)- Column density in units of cm$^{-2}$\\
Col. (4)- Value of relevant parameter of the first component. 
kT of thermal components and T$_{in}$ of the Multi-color disk model in
units of keV. The index, $\Gamma$, for the power law.\\
Col. (5)- Units of the normalization of the first component.
\emph{Power Law:}~$ph\,cm^{-2}keV^{-1}(1\,keV)$, \emph
{Raymond-Smith:}~$\frac{10^{-14}}{4 \pi (D_A(1+z))^2} \int n_e n_H dV$, \emph{Thermal Bremsstrahlung:}~$\frac{3.02\times10^{-15}}{4 \pi D^2}\int n_e n_I dV$, \emph{Multi-color disk:}~$\frac{R_{in}}{(\frac{D}{10kpc})^2}\times \cos \theta$\\
Col. (6)- Units of the second component in keV\\
Col. (7)- Units of the normalization of the second component, (see Col
5).\\
}
\end{deluxetable}
\clearpage


\begin{deluxetable}{l l l l l l l l l} 
\tablewidth{0pt}
\tabletypesize{\scriptsize}
\rotate
\tablecaption{ Values for the parameters of the fit models for the annular
region.\label{table_annulus}}
\tablehead{
\colhead{\emph{\bf Model}} & \colhead{\emph{\bf Galactic}} & \colhead{\emph{\bf Component I}} & \colhead{\emph{\bf Component I}}& \colhead{\emph{\bf Component I}} & \colhead{\emph{\bf Component II}}& \colhead{\emph{\bf Component II}} & \colhead{ $\chi^2_{\nu}$} & \colhead{\emph{\bf dof}} \\

\colhead{} & \colhead{n$_H$} & \colhead{n$_H$} &\colhead{$\Gamma$/kT} & \colhead{Norm } & \colhead{ $\Gamma$/kT} &
\colhead{Norm} \\
\colhead{(1)} & \colhead{(2)} & \colhead{(3)} & \colhead{(4)} &
\colhead{(5)} & \colhead{(6)} & \colhead{(7)}\\
}\startdata

phabs(pwlw) & & 8$\times$ 10$^{21}$  &7.2 &
1.8$\times10^{-4}$ &  &  & 7.9 & 9\\

phabs(ray) & & 9$\times$ 10$^{20}$ & 0.8 & 1.3$\times10^{-5}$ &  &  & 4.3 & 9\\

phabs(phabs*ray+pwlw) & 1.67$\times$ 10$^{20}$(frozen)  &
$5^{+3}_{-2}\times$ 10$^{21}$ & 0.26$^{+0.05}_{-0.10}$ &
$1.9^{+0.3}_{-0.5}\times 10^{-4}$ & $2.1\pm0.6$ & $4^{+2}_{-1}\times 10^{-6}$ & 1.12 & 7 \\

{\bf phabs(phabs*ray+ray) }& 1.67$\times$ 10$^{20}$(frozen) &
$4.9^{+3.0}_{-2.5}\times$ 10$^{21}$  & $0.8^{+0.1}_{-0.2}$ &
$3\pm1 \times 10^{-5}$ & $0.31^{+0.06}_{-0.05}$ & $9\pm2 \times
10^{-6}$ & 0.70 & 7\\

\enddata
\tablecomments{
Col. (1)- Spectral models: \emph{phabs:} Photoelectric Absorption,
\emph{pwlw:} Power Law, \emph{ray:} Raymond-Smith\\
Col. (2)- Column density in units of cm$^{-2}$\\
Col. (3)- Column density in units of cm$^{-2}$\\
Col. (4)- Value of relevant parameter of the first component. 
kT for thermal components in keV and the index, $\Gamma$, for the
power law.\\
Col. (5)- Units of the normalization of the first component.
\emph{Power Law:}~$ph\,cm^{-2}keV^{-1}(1\,keV)$, \emph{Raymond-Smith:}~$\frac{10^{-14}}{4 \pi (D_A(1+z))^2} \int n_e n_H dV$\\
Col. (6)- Units of the second component, (see Col 4)\\
Col. (7)- Units of the normalization of the second component, (see Col
5).\\
}
\end{deluxetable}
\clearpage

\begin{deluxetable}{l l l l l c c} 
\tablewidth{0pt}
\tabletypesize{\scriptsize}
\rotate
\tablecaption{ Unabsorbed Fluxes and luminosities.\label{table_lum}}
\tablehead{
\colhead{} & \colhead{{\bf F$_{0.23-2\,keV}$}} & \colhead{{\bf F$_{2-10\,keV}$}}
& \colhead{{\bf L$_{0.23-2\,keV}$}}
 & \colhead{{\bf L$_{2-10\,keV}$}}   & \colhead{ $\%\,_{0.23-2\,keV}$}
& \colhead{ $\%\,_{2-10\,keV}$}\\
 \colhead{} & \colhead{($10^{-14}$\,erg\,cm$^{-2}$s$^{-1}$)} &\colhead{($10^{-14} $\,erg\,cm$^{-2}$s$^{-1}$)}
 & \colhead{($10^{38} \,erg\,s^{-1}$)}           &\colhead{($10^{38}
\,erg\,s^{-1}$ )}& \colhead{} &\colhead{} 
}
\startdata
{\bf Core}          & $2.1^{+0.7}_{-1.6}$ & $2.6^{+1.0}_{-0.8}$ & $7^{+2}_{-5}$ & $8^{+3}_{-2}$    & &\\
\hspace{0.4 cm} Power Law      & $1.5\pm0.5$ & $2.6^{+1.0}_{-0.8}$ & $5\pm2$ & $8^{+3}_{-2}$  &70 & $\sim$ 100\\
\hspace{0.4 cm} Raymond-Smith  & $0.6\pm 0.2$ & $(1.2\pm0.4)\times10^{-2}(*)$ & $1.7\pm0.5$ & $(4\pm1)\times10^{-2}(*)$ &30 & 0.5\\
\tableline
{\bf Annulus }       & $11\pm2$ & $0.3\pm0.1$ & $34\pm6$   & $0.9\pm0.3$ \\
\hspace{0.4 cm} Raymond-Smith I& $2.4\pm0.5$ & $(4.1\pm0.9)\times10^{-3}(*)$& $7\pm2$ & $(1.2\pm0.3)\times10^{-2}(*)$ & 20 & 1\\
\hspace{0.3 cm}Raymond-Smith II& $9\pm3$  & $0.3\pm0.1$ & $28\pm9$  & $1.0\pm0.3$  & 80 & 99 \\
\tableline
{\bf Central $r<8''$ }         & $12^{+4}_{-3}$ &  $2.9^{+1.5}_{-1.3}$& $41^{+13}_{-30}$  & $9\pm4$    &    &   \\
\tableline
{\bf Total $r=100''$  }        &  $58^{+12}_{-10}$  & -    & $180^{+37}_{-31}$ &   -   &    & \\
\enddata
\tablecomments{The table contains the values for the luminosities and
fluxes in the energy bands 0.2-2 keV and
2-10 keV of the two regions  and the contributions of each of their
components. (*) Fluxes and luminosities estimated by extrapolation of
best fit models.}
\end{deluxetable}
\clearpage

\begin{table}
\begin{center}
\caption{Unabsorbed luminosities compared with the evolutionary model
  predictions. \label{synth_X_ray}}
\begin{tabular}{l l l  l }
\tableline \tableline
{\bf Region} & {\bf L$_{0.07-2.4\,keV}$} & {\bf L$_{2-10\,keV}$} &
Efficiency \\
 &( $10^{38}\,erg\,s^{-1}$ )&  ($10^{38}\,erg\,s^{-1}$ )&\\
\tableline
Core$\,_{observed}$(total) & $9^{+4}_{-3}$ &  $8^{+3}_{-2}$ &- \\
Core$\,_{model}$(total) & $9^{+4}_{-3}$ &  $(7.6^{+364}_{-7.4}) \times
10^{-2}$  & 0.15\\
\tableline
Core$\,_{observed}$(RS comp)& $2.0\pm 0.1$ & $(4\pm1)\times10^{-2}$&-\\
Core$\,_{model}$(RS comp) &  $2^{+0.9}_{-0.6}$ &
$(2^{+58}_{-1.8})\times10^{-2}$  & 0.03 \\ \tableline
Annulus$\,_{observed}$ & $43\pm9$ &
$0.9\pm0.3$ & -\\
Annulus$\,_{model}$ & $43^{+4}_{-3}$ & $0.34\pm0.03$  & 0.29 \\
\tableline
\end{tabular}
\tablecomments{We list the total luminosity in the core and the luminosity
  corresponding to the thermal Raymond-Smith component, as well as the
  total luminosity in the annulus, normalized to the 0.07-2.4 and 2-10
  keV bands to allow direct comparison with the model predictions. The
  error  bars  on the   model computations  correspond  to  their 90\%
  confidence  interval, associated  to  the stochastic  nature of star
  formation   \citep{cervino}.   Note   that the observed   hard X-ray
  luminosity of  the annulus has been   obtained from extrapolation of
  the thermal model fitted to the soft X-ray  band.  Last column gives
  the efficiency  in the  conversion of  mechanical  energy into X-ray
  luminosity  assumed in  the models.   The efficiency  quoted for the
  annulus correspond to an age of 3.5 Myr and $5\times 10^5$ M$_\odot$
  starburst (assuming a  Salpeter  IMF  slope  in a mass  range  1-100
  M$_\odot$ and solar metallicity). For the core region, the 4 Myr old
  starburst is $1\times   10^5$   M$_\odot$. See  the   text  for more
  details.}
\end{center}
\end{table}

\end{document}